\documentclass[showpacs,aps,pra,floatfix,reprint,superscriptaddress,footinbib,notitlepage]{revtex4-1}
\usepackage{graphicx}% Include figure files
\usepackage{amsmath}
\usepackage{amssymb} % NEW for checkmark
\usepackage{bm}% bold math
\usepackage{mathtools}
\usepackage{hyperref} \hypersetup{colorlinks=true,citecolor=blue,linkcolor=blue,urlcolor=blue}

\makeatletter
\def\pdfstartlink@attr{}
\makeatother

\makeatletter
\providecommand{\doi}[1]{%
  \begingroup
  \let\bibinfo\@secondoftwo
  \urlstyle{rm}%
  \href{http://dx.doi.org/#1}{%
    doi:\discretionary{}{}{}%
    \nolinkurl{#1}%
  }%
  \endgroup
}
\makeatother

\usepackage{listings}
\usepackage{multirow}
\usepackage{booktabs} % to use rule
\AtBeginDocument{ % to use booktabs in reftex
\heavyrulewidth=.08em
\lightrulewidth=.05em
\cmidrulewidth=.03em
\belowrulesep=.65ex
\belowbottomsep=0pt
\aboverulesep=.4ex
\abovetopsep=0pt
\cmidrulesep=\doublerulesep
\cmidrulekern=.5em
\defaultaddspace=.5em
}
\usepackage[dvipsnames]{xcolor}
\usepackage{soul}
\usepackage{makecell} %DX uncommented out
\usepackage{tabularx}
\newcolumntype{Y}{>{\raggedleft\arraybackslash}X}
\usepackage{tabu} % NEW
\usepackage{longtable} % NEW
\usepackage{supertabular} % NEW
\usepackage{enumitem}
\usepackage{xcolor} % to color verb environment
\usepackage{newverbs} % to color verb environment
\usepackage{mhchem} % format chemical equations
\newverbcommand{\cverb}{}{} % to color verb environment
\newcommand{\verbWrap}{\lstinline[basicstyle=\normalsize\ttfamily,postbreak=]}
\newcommand{\verbWrapBlue}{\lstinline[basicstyle=\normalsize\ttfamily,postbreak=]}

\DeclareFixedFont{\ttb}{T1}{txtt}{bx}{n}{8} % for bold
\DeclareFixedFont{\ttm}{T1}{txtt}{m}{n}{8}  % for normal

\definecolor{pranab_green}{rgb}{0.31,0.53,0.10}
\definecolor{pranab_red}{rgb}{0.85,0.23,0.11}
\definecolor{deepblue}{rgb}{0,0,0.5}
\definecolor{deepred}{rgb}{0.6,0,0}
\definecolor{deepgreen}{rgb}{0,0.5,0}
\definecolor{tmrwBlue}{rgb}{0.259,0.443,0.68.2}
\definecolor{tmrwRed}{rgb}{0.784,0.157,0.161}
\definecolor{tmrwGreen}{rgb}{0.443,0.549,0}
\definecolor{tmrwPurple}{rgb}{0.537,0.349,0.659}
\definecolor{tmrwAqua}{rgb}{0.243,0.6,0.624}
\definecolor{tmrwYellow}{rgb}{0.918,0.718,0}
\definecolor{tmrwOrange}{rgb}{0.871,0.576,0.373}
\definecolor{tmrwComment}{rgb}{0.557,0.565,0.549}

\newcommand\pythonstyle{\lstset{
language=Python,
basicstyle=\ttm,
otherkeywords={__init__, self},             % Add keywords here
keywordstyle=\ttb\color{tmrwBlue},
emph={and,break,class,continue,def,yield,del,elif ,else,%
except,exec,finally,for,from,global,if,import,as,%
lambda,not,or,pass,print,raise,return,try,while,assert,with},
emphstyle=\ttb\color{tmrwPurple},    % Custom highlighting style
emph={[2]},
emphstyle=[2]\ttb\color{tmrwBlue},
emph={[3] in },
emphstyle=[3]\ttb\color{tmrwAqua},
emph={[4]object,type,list,set,len,dict,tuple,str,repr,int,float},
emphstyle=[4]\ttb\color{tmrwYellow},
emph={[5]aflow_sym, pprint, Symmetry, json, subprocess, os, aflow_command, get_symmetry, get_edata, get_sgdata, loads, path, realpath, XtalFinder, compare_materials, compare_materials_directory, compare_materials_file, compare_structures, compare_structures_directory, compare_structures_file, compare2database, compare2prototypes, get_prototype_label, get_unique_permutations, get_unique_atom_decorations, isopointal_prototypes},
emphstyle=[5]\ttm\color{tmrwBlue},
emph={[6]True, False, None},
emphstyle=[6]\ttm\color{tmrwOrange},
emph={[7]self},
emphstyle=[7]\ttm\color{tmrwRed},
stringstyle=\color{tmrwGreen},
morecomment=[s]{"""}{"""},
commentstyle=\color{tmrwComment}\ttm,
literate=
 {-}{{{-}}}1
 {<}{{{<}}}1,
frame=tb,                         % Any extra options here
breaklines=true,
postbreak=\mbox{\textcolor{tmrwRed}{$\hookrightarrow$}\space},
showstringspaces=false            %
}}

\lstnewenvironment{python}[1][]
{
  \pythonstyle
  \lstset{#1}
}
{}

\newcommand\pythoninline[1]{{\pythonstyle\lstinline!#1!}}

\lstdefinelanguage{mylang}{
  basicstyle=\ttfamily,
  alsoletter=0123456789,
  alsodigit={.-}
}
\setlist[itemize]{noitemsep, topsep=0pt}
\newlist{myitemize}{itemize}{3}
\setlist[myitemize,1]{label=\textbullet,leftmargin=1em}
\setlist[myitemize,2]{label=--,leftmargin=1em}
\setlist[myitemize,3]{label=$\diamond$,leftmargin=1em}
\setlist[myitemize]{noitemsep, topsep=0pt}

\setlist[itemize]{noitemsep, topsep=0pt}
\newlist{myitemize_nobullet}{itemize}{3}
\setlist[myitemize_nobullet,1]{label=,leftmargin=1em}
\setlist[myitemize_nobullet,2]{label=,leftmargin=1em}
\setlist[myitemize_nobullet,3]{label=,leftmargin=1em}
\setlist[myitemize_nobullet]{noitemsep, topsep=0pt}

\bibpunct{[}{]}{,}{n}{}{} %square brackets for refs

\setstcolor{red} %correction
\let\oldfootnote\footnote
\def\footnote{\ifhmode\unskip\fi\oldfootnote}

\def\AFLOW{{\small AFLOW}}
\def\AFLOWVERSION{{{3.2}}} 
\def\STRUCTUREMATCHERVERSION{{{2020.4.2}}} 
\def\XTALCOMPVERSION{{{downloaded from GitHub on 14 Apr. 2020}}} 
\def\AFLOWICSDDATEXTALMATCH{14 August 2020} %2 August 2019, 20 June, %2019, 14 August 2020
\def\NUMBEROFENTHALPYCOMPARISONS{6,795} %6,813 (20 June 2019)
\def\AFLOWXTALFINDERAFLOWVERSION{3.2} % version of AFLOW containing AFLOW-XTAL-MATCH functionality
\def\SUPPLEMENTARYMATERIALS{Supplementary Information} % easy fix for "Supplementary" vs "Supporting" and "Materials" vs "Information" 
\def\MISFITAUTHORS{Burzlaff and Malinovsky} % easy fixes 
\def\MISFITAUTHORSPOSSESIVE{Burzlaff and Malinovsky's} % easy fixes 
\def\QUANTUMESPRESSO{\textsc{Quantum {\small ESPRESSO}}}
\def\FHIAIMS{{\small FHI-AIMS}}
\def\ABINIT{{\small ABINIT}}
\def\AFLUX{{\small AFLUX}}
\def\AURL{{\small AURL}}
\def\RESTAPI{{\small REST-API}}
\def\REST{{\small REST}}
\def\API{{\small API}}
\def\APIS{{\small API}s}
\def\ELK{{\small ELK}}

\def\VASP{{\small VASP}}
\def\INCAR{{\small INCAR}}
\def\OUTCAR{{\small OUTCAR}}
\def\ICSD{{\small ICSD}}
\def\PROTOTYPEENCYCLOPEDIA{{Prototype Encyclopedia}} %LibProto

\def\FDAT{{\small FDAT}}
\def\CSD{{\small CSD}}
\def\OQMD{{\small OQMD}}

\def\CIF{{\small CIF}}
\def\CIFTWOCELL{{\small CIF2Cell}}

\def\POSCAR{{\small POSCAR}}

\def\AIIDA{{\small AiiDA}}
\def\AFLOWXTALFINDER{{\small AFLOW}-XtalFinder}
\def\AFLOWXTALFINDERSHORT{{XtalFinder}}
\def\AFLOWSYM{{\small AFLOW-SYM}}
\def\AFLOWCHULL{{\small AFLOW-CHULL}}

\def\XTALOPT{{\small XTALOPT}}
\def\CALYPSO{{\small CALYPSO}}
\def\KPLOT{{\small KPLOT}}
\def\FDAT{{\small FDAT}}
\def\CSD{{\small CSD}}
\def\ITC{{\small ITC}}
\def\LDAU{{\small LDAU}}

\def\PLATON{{Platon}}

\def\CRYCOM{{\small CRYCOM}}
\def\STRUCTURETIDY{{\small STRUCTURE-TIDY}}
\def\XTALCOMP{{\small XTALCOMP}}
\def\STRUCTUREMATCHER{{Structure Matcher}}
\def\SPAP{{\small SPAP}}
\def\COMPSTRU{{\small COMPSTRU}}
\def\CMPZ{{\small CMPZ}}

\def\NOMAD{{NoMaD}}
\def\LFA{{\small LFA}}
\def\JSON{{\small JSON}}
\def\GNU{{\small GNU}}
\def\GPL{{\small GPL}}
\def\LIBONE{{\small LIB1}}
\def\LIBTWO{{\small LIB2}}
\def\LIBTHREE{{\small LIB3}}
\def\IDEALPROTOTYPESECTION{{Problem of the ideal prototype}}
\def\SYMBOLICCPLUSPLUS{{\small SymbolicC++}}
\usepackage{pifont} % for xmark
\def\AFLOWXMARK{x}

\def\citeANRL{\cite{curtarolo:art121,curtarolo:art145}}

\makeatletter \renewcommand\frontmatter@abstractwidth{\dimexpr\textwidth\relax} \makeatother  % WIDEABSTRACT
\begin{document}

%\title{\Large AFLOW-XtalFinder: Automated framework for quantifying the structural similarity of materials and identifying unique crystal prototypes}
%\title{\large AFLOW-XtalFinder: a reliable choice to identify similarities in crystalline materials}
\title{\large AFLOW-XtalFinder: a reliable choice to identify crystalline prototypes}

\author{David Hicks}
\affiliation{Department of Mechanical Engineering and Materials Science, Duke University, Durham, North Carolina 27708, USA}
\affiliation{Center for Autonomous Materials Design, Duke University, Durham, North Carolina 27708, USA}
\author{Cormac Toher}
\affiliation{Department of Mechanical Engineering and Materials Science, Duke University, Durham, North Carolina 27708, USA}
\affiliation{Center for Autonomous Materials Design, Duke University, Durham, North Carolina 27708, USA}
\author{Denise C. Ford}
\affiliation{Department of Mechanical Engineering and Materials Science, Duke University, Durham, North Carolina 27708, USA}
\affiliation{Center for Autonomous Materials Design, Duke University, Durham, North Carolina 27708, USA}
\author{Frisco Rose}
\affiliation{Department of Mechanical Engineering and Materials Science, Duke University, Durham, North Carolina 27708, USA}
\affiliation{Center for Autonomous Materials Design, Duke University, Durham, North Carolina 27708, USA}
\author{\\Carlo De Santo}
\affiliation{Department of Mechanical Engineering and Materials Science, Duke University, Durham, North Carolina 27708, USA}
\affiliation{Center for Autonomous Materials Design, Duke University, Durham, North Carolina 27708, USA}
\author{Ohad Levy}
\affiliation{Department of Mechanical Engineering and Materials Science, Duke University, Durham, North Carolina 27708, USA}
\affiliation{Center for Autonomous Materials Design, Duke University, Durham, North Carolina 27708, USA}
\affiliation{Department of Physics, NRCN, P.O. Box 9001, Beer-Sheva 84190, Israel}
\author{Michael J. Mehl}
%\affiliation{United States Naval Academy, Annapolis, Maryland 21402, USA}
\affiliation{Department of Mechanical Engineering and Materials Science, Duke University, Durham, North Carolina 27708, USA}
\affiliation{Center for Autonomous Materials Design, Duke University, Durham, North Carolina 27708, USA}
\author{Stefano Curtarolo}
\email[]{stefano@duke.edu}
\affiliation{Department of Mechanical Engineering and Materials Science, Duke University, Durham, North Carolina 27708, USA}
\affiliation{Center for Autonomous Materials Design, Duke University, Durham, North Carolina 27708, USA}

%%%%%%%%%%%%%%%%%%%%%%%%%%%%%%%%%%%%%%%%%%%%%%%%%%%%%%%%%%%%%%%%%%%%% 
%% date
%%%%%%%%%%%%%%%%%%%%%%%%%%%%%%%%%%%%%%%%%%%%%%%%%%%%%%%%%%%%%%%%%%%%% 
\date{\today}% It is always \today, today,

%%%%%%%%%%%%%%%%%%%%%%%%%%%%%%%%%%%%%%%%%%%%%%%%%%%%%%%%%%%%%%%%%%%%%%%%%%%%%%%%%%%%%%%%%%%%%%%%%%%%%%%%%%%%%%%%%%%%%% 
% 
% START: Abstract
% 
%%%%%%%%%%%%%%%%%%%%%%%%%%%%%%%%%%%%%%%%%%%%%%%%%%%%%%%%%%%%%%%%%%%%%%%%%%%%%%%%%%%%%%%%%%%%%%%%%%%%%%%%%%%%%%%%%%%%%% 
\begin{abstract}
  \noindent The accelerated growth rate of repository entries in crystallographic databases 
  makes it arduous to identify and classify their prototype structures. 
  The {{open-source {\small AFLOW}-XtalFinder package}} was developed to solve this problem. 
  It {{symbolically}} maps structures into standard
  designations {{following the {\small AFLOW} Prototype
  Encyclopedia}} and calculates the internal degrees of freedom consistent with the 
  International Tables for Crystallography.
  {{To ensure uniqueness}}, structures are analyzed and compared via
  symmetry, local atomic geometries, and crystal mapping techniques,
  {{simultaneously grouping them by similarity.}}
  The software
  \textbf{i.} distinguishes distinct crystal prototypes and atom decorations,
  \textbf{ii.} determines equivalent spin configurations, 
  \textbf{iii.} reveals compounds with similar properties, and 
  \textbf{iv.} guides the discovery of unexplored materials. 
  {{The operations are accessible through a Python module ready for  workflows, and through command line syntax.}}
   {{All the 4+ million compounds in the {\small AFLOW}.org repositories are mapped to their
  ideal prototype, allowing users to search database entries via
  symbolic structure-type.
  Furthermore, 15,000 unique structures --- sorted by prevalence --- are extracted from the 
  {\small AFLOW}-{\small ICSD} catalog to serve as future prototypes
  in the Encyclopedia.}}
\end{abstract}
\maketitle

%%%%%%%%%%%%%%%%%%%%%%%%%%%%%%%%%%%%%%%%%%%%%%%%%%%%%%%%%%%%%%%%%%%%%%%%%%%%%%%%%%%%%%%%%%%%%%%%%%%%%%%%%%%%%%%%%%%%%% 
% 
% END: Abstract
% 
%%%%%%%%%%%%%%%%%%%%%%%%%%%%%%%%%%%%%%%%%%%%%%%%%%%%%%%%%%%%%%%%%%%%%%%%%%%%%%%%%%%%%%%%%%%%%%%%%%%%%%%%%%%%%%%%%%%%%% 

% ELSEARTICLE \begin{keyword}
% ELSEARTICLE   Crystal comparison \sep structure prototypes \sep high-throughput \sep materials genomics 
% ELSEARTICLE   \sep AFLOW \sep AFLOW-XTAL-MATCH

%   PACS codes here, in the form: \PACS code \sep code 
%   MSC codes here, in the form: \MSC code \sep code
%   or \MSC[2008] code \sep code (2000 is the default)
%   ELSEARTICLE \end{keyword}

% ELSEARTICLE \end{frontmatter}


{{
Scientists have been struggling for decades to identify prototypes 
({\it{e.g.}} {\it{Strukturbericht}} series~\cite{Ewald_Struk_I_1931} and Pearson's Handbook~\cite{Villars91})
and duplicates in crystallographic databases; and to label 
structures in a concise way to recognize (and enable searching by) structure-types.
The recent rapid growth of online repositories has worsened the problem \cite{nmatHT}.}}
%and paradoxically, the recent rapid growth of online repositories has worsened the problem \cite{nmatHT}.
%Scientists have been struggling to identify prototypes 
%({\it{e.g.}} {\it{Strukturbericht}} series~\cite{Ewald_Struk_I_1931} and Pearson's Handbook~\cite{Villars91})
%and duplicates in crystallographic databases for decades, 
%and paradoxically, the recent rapid growth of online repositories has worsened the problem \cite{nmatHT}.
%Recognizing distinct crystalline compounds is becoming increasingly difficult, 
{{Distinguishing}} distinct crystalline compounds is becoming increasingly difficult, 
leading to repetition of previously studied materials,
hindering database variety 
--- biasing data-driven analyses and machine learning methods~\cite{Kolcz_MLDataDuplication_2003,Muratov_et_al_QSAR_chemsocrev_2020} ---  
and wasting valuable computational and experimental resources.
The multitude of crystal geometries make by-hand detection of prototypes and repeated entries intractable.
A major complication for finding structure-types is the non-standard representation 
of crystals.  
Determination of unique crystallographic structures is obfuscated by 
\textbf{i.} unit cell representations and 
\textbf{ii.} origin choices.   
While standard forms exist --- such as Niggli~\cite{Niggli1928} and Minkowski~\cite{MinkowskiReduction}
unit cells --- the conversion procedures are highly sensitive to numerical tolerance 
values and can cast similar structures into differing descriptions~\cite{crycom_1994,xtal-comp_2012}.  
Additionally, lattice standardization techniques do not address differences 
in origin choices.  
The lack of commensurate representations impedes the search for prototypes and 
inhibits mappings between similar crystals and their corresponding properties. %, obscuring the structure-property relationship.  

To overcome non-standard descriptions, crystal comparison tools have been 
developed to identify similar structures.  
Programs such as 
\STRUCTUREMATCHER~\cite{pymatgen-structure-matcher_2011},
\XTALCOMP~\cite{xtal-comp_2012},
\SPAP~\cite{spap_jpcm_2017},
\CMPZ~\cite{Hundt_CMPZ_2006},
\CRYCOM~\cite{crycom_1994},
\STRUCTURETIDY~\cite{structure-tidy_1987}, and
\COMPSTRU~\cite{Flor_COMPSTRU_2016}
are available with varying objectives related to structure comparison.
For instance, \XTALCOMP\ is coupled with the \XTALOPT\ infrastructure for 
identifying distinct materials generated 
via their evolutionary algorithm~\cite{xtal-opt_2011}.
Despite the considerable number of platforms, none are suitable for autonomous prototype detection.
Crystallographic symmetry is neglected in \STRUCTUREMATCHER, \XTALCOMP, and \SPAP; 
while \STRUCTURETIDY, \CRYCOM, and \COMPSTRU\ rely on external symmetry packages.
Additionally, most tools only feature single pairwise comparisons (with the exception of \STRUCTUREMATCHER) and 
others require additional inputs ({\it{e.g.}} space group, Wyckoff positions, and unit cell choice).
Aside from technical functionality, the codes do not offer built-in methods 
to compare structures to existing crystallographic libraries and material repositories.
To promote materials discovery, 
routines must analyze compounds with respect to established prototypes to identify new structure-types.
This would enable the expansion of prototype libraries --- such 
as the \AFLOW\ \PROTOTYPEENCYCLOPEDIA\ (or \PROTOTYPEENCYCLOPEDIA\ for brevity)~\citeANRL{} --- 
fueling generation of unique compounds via prototype decoration.
Comparing compounds to those in materials databases can prevent duplication.
Moreover, the properties of database entries can be used to estimate those of similar uncalculated compounds, 
exploiting the structure-property relationship of materials.
%These shortcomings hinder rapid and effective structural analyses and incorporation into programmatic workflows.
Clearly, an automatic and reliable large-scale method for discerning unique crystallographic structures is 
therefore crucial for the materials science community.

\begin{figure*}[]
  \centering
  \includegraphics[width=1\textwidth]{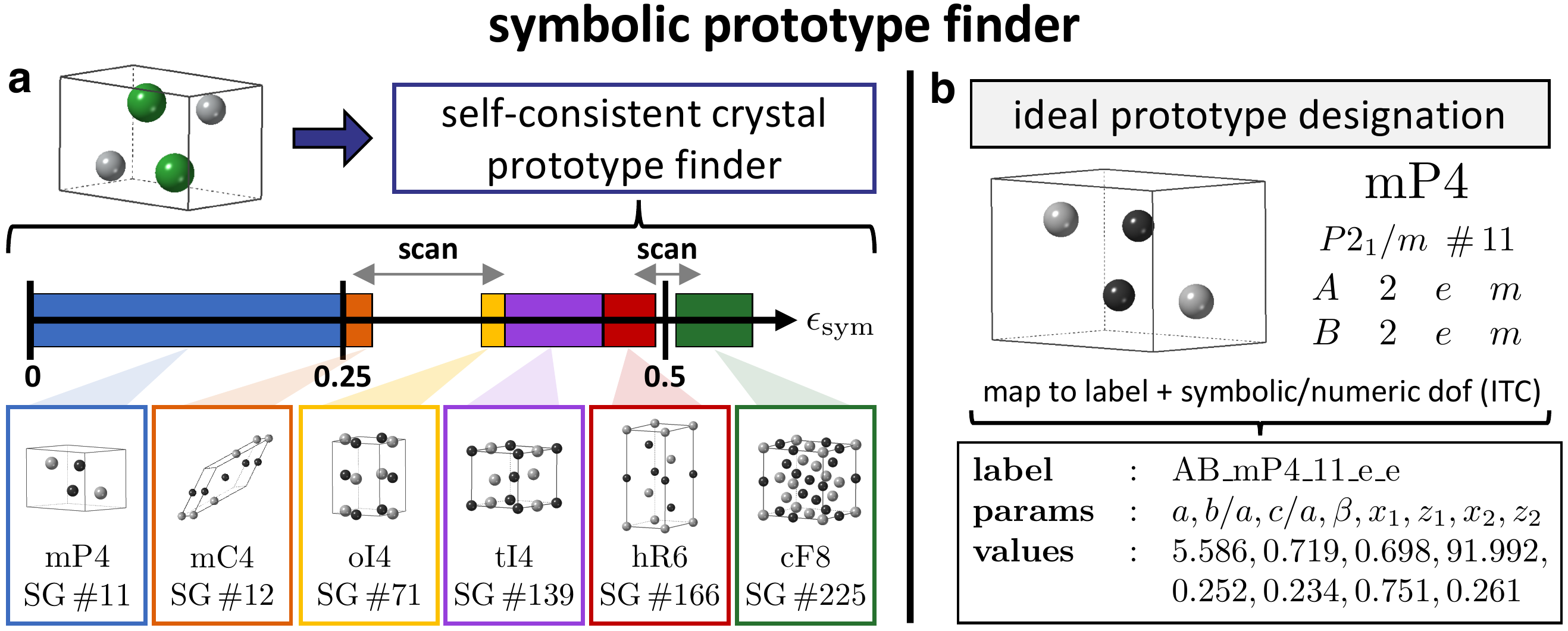}
    \caption{{\textbf{Self-consistent symbolic prototype finder.}}
    \small
    {{
    (\textbf{a}) The prototype for an input structure, {\it{e.g.}} AlCl (\ICSD\ \#56541), is identified by
                 analyzing its symmetry.  
                 Classification of the prototype may change depending on the symmetry tolerance ($\epsilon_{\mathrm{sym}}$):
                 mP4, SG \#11 ($0 < \epsilon_{\mathrm{sym}} \le 0.25~\mathrm{\AA}$);
                 mC4, SG \#12 ($0.26 \le \epsilon_{\mathrm{sym}} \le 0.27~\mathrm{\AA}$);
                 oI4, SG \#71 ($0.34 \le \epsilon_{\mathrm{sym}} \le 0.36~\mathrm{\AA}$);
                 tI4, SG \#139 ($0.37 \le \epsilon_{\mathrm{sym}} \le 0.44~\mathrm{\AA}$);
                 hR6, SG \#166 ($0.45 \le \epsilon_{\mathrm{sym}} \le 0.49~\mathrm{\AA}$); and
                 cF8, SG \#225 ($0.51 \le \epsilon_{\mathrm{sym}} \le 1.0~\mathrm{\AA}$).
                 An adaptive routine is employed for tolerance regions with incommensurate symmetry descriptions 
                 (gray arrows for 
                 $0.27 < \epsilon_{\mathrm{sym}} < 0.34~\mathrm{\AA}$ and $0.49 < \epsilon_{\mathrm{sym}} < 0.51~\mathrm{\AA}$), 
                 ensuring self-consistent prototype/symmetry designations.
    (\textbf{b}) The structure is then mapped into its prototype label and symbolic and numeric internal 
                 \underline{d}egrees \underline{o}f \underline{f}reedom (dof), consistent with the 
                 \underline{I}nternational \underline{T}ables for \underline{C}rystallography (\ITC).
                 Structures in this representation can be generated with the symbolic prototype generator.
    }} % \textcolor
    }
  \label{fig1_label}
\end{figure*}

\AFLOWXTALFINDER\ (\underline{AFLOW} crys\underline{tal} \underline{finder}, \AFLOWXTALFINDERSHORT\ for brevity) addresses many of the previously mentioned issues in a high-throughput fashion.
{{The primary objective of \AFLOWXTALFINDERSHORT\ is to
identify/classify the prototypes of materials and relate them via structural similarity metrics.
To accomplish this, \AFLOWXTALFINDERSHORT\ determines the ideal 
prototype designation of crystal structures, consistent with the 
\underline{I}nternational \underline{T}ables for \underline{C}rystallography (\ITC)~\cite{tables_crystallography}.
Any structure in this representation can be automatically generated via
a new symbolic prototype generator.
Similarity between structures is analyzed on multiple fronts.}}
Crystallographic structures are first compared by symmetry (isopointal analysis),
leveraging a robust software implementation, \AFLOWSYM, which calculates
self-consistent symmetry descriptions 
freeing the user from tolerance adjustments~\cite{curtarolo:art135}.
Local atomic geometries are also computed to match neighborhoods of atoms in crystals (isoconfigurational snapshots).
Finally, crystal similarity is resolved by rigorous structure mapping procedures (complete isoconfigurational analysis) 
and quantified via a misfit criterion~\cite{Burzlaff_ActaCrystA_1997}.
The prototype finder accommodates automatic workflows, with functionality to 
analyze multiple materials/structures simultaneously via multithreading.  
Features are provided to identify crystallographic structures, distinct materials, 
atom decorations, and spin configurations. 
Methods are also included to compare compounds{{/prototypes} to the \AFLOW.org repository and \AFLOW\ prototype libraries.
{{Every entry in the \AFLOW.org repository has been mapped to its prototype label,
enabling users to search the database by structure-type.
%Lastly, procedures are introduced to cast structures into the \AFLOW\ standard prototype designation, 
%streamlining prototype design. %and exploration of new materials via atom decoration.
The \AFLOWXTALFINDERSHORT\ code} --- written in C++ --- is part of the \AFLOW\ (\underline{A}utomatic \underline{flow})
framework~\cite{aflow_fleet_chapter_full,aflowAEL,aflowAGL,curtarolo:art65} and is open-source under the \GNU-\GPL\ license.  
For seamless integration into different work environments, this functionality is accessible via the command-line 
and a Python module.
%\AFLOWXTALFINDERSHORT\ is a resource to pinpoint uncharted
%domains of crystallographic space and expedite the design of innovative materials.

\label{subsec:prototyping}
\noindent{{\textbf{\IDEALPROTOTYPESECTION}.}}
{{
Prototype structures are generally classified in terms of their symmetry characteristics.
For example, the rocksalt prototype has 
a face-centered cubic lattice and 8 atoms in the conventional cell ({\it{i.e.}} Pearson symbol of cF8), 
space group $Fm\bar{3}m$ (\#225), and 
Wyckoff positions $4 \, \, a \, \,m\bar{3}m$ and $4 \, \, b \, \, m\bar{3}m$.
Determining this information for any arbitrary structure is often 
a challenge: numerical noise in the atomic positions inhibits
detection of crystal isometries, requiring by-hand modification of tolerance thresholds.
%, and is intensified for lower-symmetry compounds.
Furthermore, consistency between real- and reciprocal-space symmetries
is often overlooked, and yet it is imperative for reliable {\it{ab initio}} simulations. 
Thus, accurate prototype detection relies on robust symmetry analyses.
}}

{{
\AFLOWXTALFINDERSHORT\ employs a self-consistent mechanism
to find the ideal prototype of a given structure.
The space group, Pearson symbol, and occupied Wyckoff positions
are calculated via the \AFLOWSYM\ routines~\cite{curtarolo:art135}.
The prototype classification is sensitive to the symmetry tolerance ($\epsilon_{\mathrm{sym}}$).
For example, the AlCl structure (\ICSD\ \#56541, DFT-relaxed) in Figure~\ref{fig1_label}(a) 
can be classified as one of six different prototypes as a function of $\epsilon_{\mathrm{sym}}$: 
\textbf{i.}~mP4, SG \#11 ($0 < \epsilon_{\mathrm{sym}} \le 0.25~\mathrm{\AA}$);
\textbf{ii.}~mC4, SG \#12 ($0.26 \le \epsilon_{\mathrm{sym}} \le 0.27~\mathrm{\AA}$);
\textbf{iii.}~oI4, SG \#71 ($0.34 \le \epsilon_{\mathrm{sym}} \le 0.36~\mathrm{\AA}$);
\textbf{iv.}~tI4, SG \#139 ($0.37 \le \epsilon_{\mathrm{sym}} \le 0.44~\mathrm{\AA}$);
\textbf{v.}~hR6, SG \#166 ($0.45 \le \epsilon_{\mathrm{sym}} \le 0.49~\mathrm{\AA}$); and
\textbf{vi.}~cF8, SG \#225 ($0.51 \le \epsilon_{\mathrm{sym}} \le 1.0~\mathrm{\AA}$).
For certain tolerance values 
--- {\it{e.g.}} $0.27 < \epsilon_{\mathrm{sym}} < 0.34~\mathrm{\AA}$ and
$0.49 < \epsilon_{\mathrm{sym}} < 0.51~\mathrm{\AA}$ ---
incommensurate symmetry descriptions are calculated. 
To overcome this, the symmetry tolerance is automatically changed, 
scanning tighter and looser tolerances around the initial value,
to find consistent symmetry descriptions at a new $\epsilon_{\mathrm{sym}}$.
This autonomous approach ensures prototype classifications are 
correct and compatible against all symmetry descriptors ({\it{e.g.}} space group,
Wyckoff position, lattice type, Brillouin zone, etc.).
}}

{{
The default symmetry tolerance value for classifying prototypes 
in \AFLOWXTALFINDERSHORT\ is proportional to the minimum interatomic distance 
($d^{\mathrm{min}}_{\mathrm{nn}}$/100).
The tolerance is thus system-specific, and it has been shown to be consistent with experimentally
resolved symmetries~\cite{curtarolo:art135}.
Nevertheless, the tolerance can also be adjusted by the user, and is guaranteed to return
a commensurate designation due to the adaptive prototype protocol shown in Figure~\ref{fig1_label}(a).
}}
{{Once the symmetry attributes of the crystal are calculated, 
\AFLOWXTALFINDERSHORT\ automatically 
maps the structure to its \AFLOW\ prototype label and symmetry-based degrees of freedom
(Figure~\ref{fig1_label}(b)),
{\it{i.e.}} lattice parameters/angles and non-fixed Wyckoff coordinates~\citeANRL.
These designations are commensurate with the \ITC\ cell choices and Wyckoff positions; the {\it{de facto}} standard
for crystallography.
The label specifies the stoichiometry and symmetry of the structure in underscore-separated fields. 
The fields indicate the following (example system: esseneite structure, ABC6D2\_mC40\_15\_e\_e\_3f\_f~\cite{curtarolo:art121}) 
\begin{itemize}
    \item{first field: the reduced stoichiometry based on alphabetic ordering of the compound, {\it{e.g.}} a quaternary with stoichiometry ABC6D2,}
    \item{second field: the Pearson symbol, {\it{e.g.}} mC40,}
    \item{third field: the space group number, {\it{e.g.}} space group \#15,}
    \item{fourth field: the Wyckoff letter(s) of the first atomic site, {\it{e.g.}} site $A$: one Wyckoff position with letter $e$,}
    \item{fifth field: the Wyckoff letter(s) of the second atomic site, {\it{e.g.}} site $B$: one Wyckoff position with letter $e$,}
    \item{sixth field: the Wyckoff letter(s) of the third atomic site, {\it{e.g.}} site $C$: three Wyckoff positions with letters $f$, and}
    \item{seventh field: the Wyckoff letter(s) of the fourth atomic site, {\it{e.g.}} site $D$: one Wyckoff position with letter $f$.}
\end{itemize}
}}
{{
The prototype parameters specify the degrees of freedom allowed by the symmetry of the structure.
For the esseneite structure, there are 18 parameters:
$a$, $b/a$, $c/a$, $\beta$, $y_{1}$, $y_{2}$, $x_{3}$, $y_{3}$, $z_{3}$, $x_{4}$, $y_{4}$, $z_{4}$, $x_{5}$, $y_{5}$, $z_{5}$, $x_{6}$, $y_{6}$, and $z_{6}$.
The first three variables are the lattice parameters --- with $b$ and $c$ represented in relation to $a$ ---
the fourth variable is the lattice angle $\beta$, and the subsequent variables are the 
Wyckoff coordinates (fractional) that are not fixed by symmetry. % confined -> fixed
The sequence of the Wyckoff parameters is based on the alphabetic ordering of the Wyckoff letters,
followed by alphabetic ordering of the species.
Additional information regarding the label and parameters are discussed in the Refs.~\citeANRL.
}}

{{
Mapping structures into this format characterizes prototypes in 
a concise and descriptive manner.
The representation also easily distinguishes isopointal and isoconfigurational prototypes.
Two compounds with similar labels are isopointal ({\it{i.e.}} same symmetry), and 
are isoconfigurational if their parameters are the same ({\it{i.e.}} equivalent geometric configurations){ }~\footnote{
    A strict parameter comparison does not distinguish isoconfigurational structures, 
    {\it{e.g.}} parameters may differ by an origin shift.
}.
Moreover, the representation reveals the degrees of freedom that can be altered, 
while preserving the underlying symmetry.  
This is useful for showing continuous structure transitions within the same symmetry-type 
and performing symmetry-constrained structure relaxations~\cite{Lenz_SymConstrain_2019}.
Lastly, with this format, structures are now easily regenerated with the \AFLOW\ software.
}}

\label{subsec:prototype_generator}
\noindent{{\textbf{Symbolic prototype generator}.}}
{{
Structures represented in the ideal prototype designation can be
created and decorated with any atomic elements via a new symbolic prototype generator,
enabling automatic materials design.
A procedure --- introduced in Refs.~\citeANRL --- 
has been extended to create all possible prototype structures,
going beyond those previously described in the \PROTOTYPEENCYCLOPEDIA.
Given a crystal's composition, Pearson symbol, space group, and occupied Wyckoff positions, 
the generator determines the degrees of freedom in symbolic notation 
({\it{i.e.}} $a$, $b$/$a$, $c$/$a$, $\alpha$, $\beta$, $\gamma$, $x$, $y$, and $z$) 
that must be specified, based on the \ITC\ conventions~\cite{tables_crystallography}.
Feeding in the ideal prototype label and degrees of freedom to the symbolic 
generator will produce the corresponding geometry file,
substituting the appropriate degrees of freedom with the input values.
Prototypes, including those in the \PROTOTYPEENCYCLOPEDIA, no longer need to be
tabulated (hard-coded) in the \AFLOW\ software, and are now created on-the-fly.
With this prototype generator, \AFLOW\ is capable of creating structures to span all regions of crystallographic space.
}}

{{
    Structures are generated with the following
    prototype command syntax: 
    \verbWrap!--proto=label!  
    \verbWrap!--params=parameter_1,parameter_2,...!.
    Here, the \verbWrap!label! is the ideal prototype label,
    {\it{e.g.}} AB\_mP4\_11\_e\_e as shown in Figure~\ref{fig1_label}({b}), and
    \verbWrap!parameter_1,parameter_2,...! are the comma-separated values for the prototype's degrees of freedom,
    {\it{e.g.}} 5.586, 0.719, 0.698, 91.992, 0.252, 0.234, 0.751, and 0.261 as shown in Figure~\ref{fig1_label}({b}).
    By default, structures are generated with fictitious species in alphabetical order ({\it{i.e.}} A, B, C, D, etc.).
    Users can override this order by specifying other permutations after the prototype label (separated by a period), 
    {\it{i.e.}} \verbWrap!--proto=label.BAC...!; a useful feature for controlling the atomic site decorations.
    Specific elements can be decorated onto the prototype by appending the element abbreviations to the command
    in colon-separated alphabetical order, {\it{e.g.}} \verbWrap!--proto=label:Ag:Cu:Zr!.
    The generator checks for any inconsistencies with the provided label and/or parameter values, 
    terminating prematurely with a message listing possible fixes to the command.
    The generator supports multiple geometry file formats, including
    \VASP\ (\POSCAR)~\cite{kresse_vasp}, \FHIAIMS~\cite{Blum_CPC2009_AIM}, 
    \QUANTUMESPRESSO~\cite{quantum_espresso_2009}, 
    \ABINIT~\cite{gonze:abinit},  \ELK~\cite{Elk_LAPW}, and \CIF.
    Swapping the command \verbWrap!--proto=label! with \verbWrap!--aflow_proto=label!,
    will build an aflow.in file, \AFLOW's input file (using a standard set of DFT parameters by default~\cite{curtarolo:art104}),
    automating {\it{ab initio}} simulations of these compounds.
  }}

{{
    The generator can also print the symbolic representation of the
    lattice and Wyckoff positions.
Adding the option
\verbWrap!--add_equations! to the prototype command
returns both a numerical and symbolic version of the geometry file, and 
the option
\verbWrap!--equations_only! only prints the symbolic version.
Symbolic geometry files can be printed with 
respect to the conventional cell (\ITC) or
symbolically transformed into the primitive cell 
(using the \SYMBOLICCPLUSPLUS\ open-source software~\cite{Hardy_SymbolicCPlusPlus_2008}).
By default, \AFLOW\ provides the primitive cell, since fewer-atom unit cells
are more computationally efficient.
}}

\begin{figure*}[]
  \centering
  \includegraphics[width=1\textwidth]{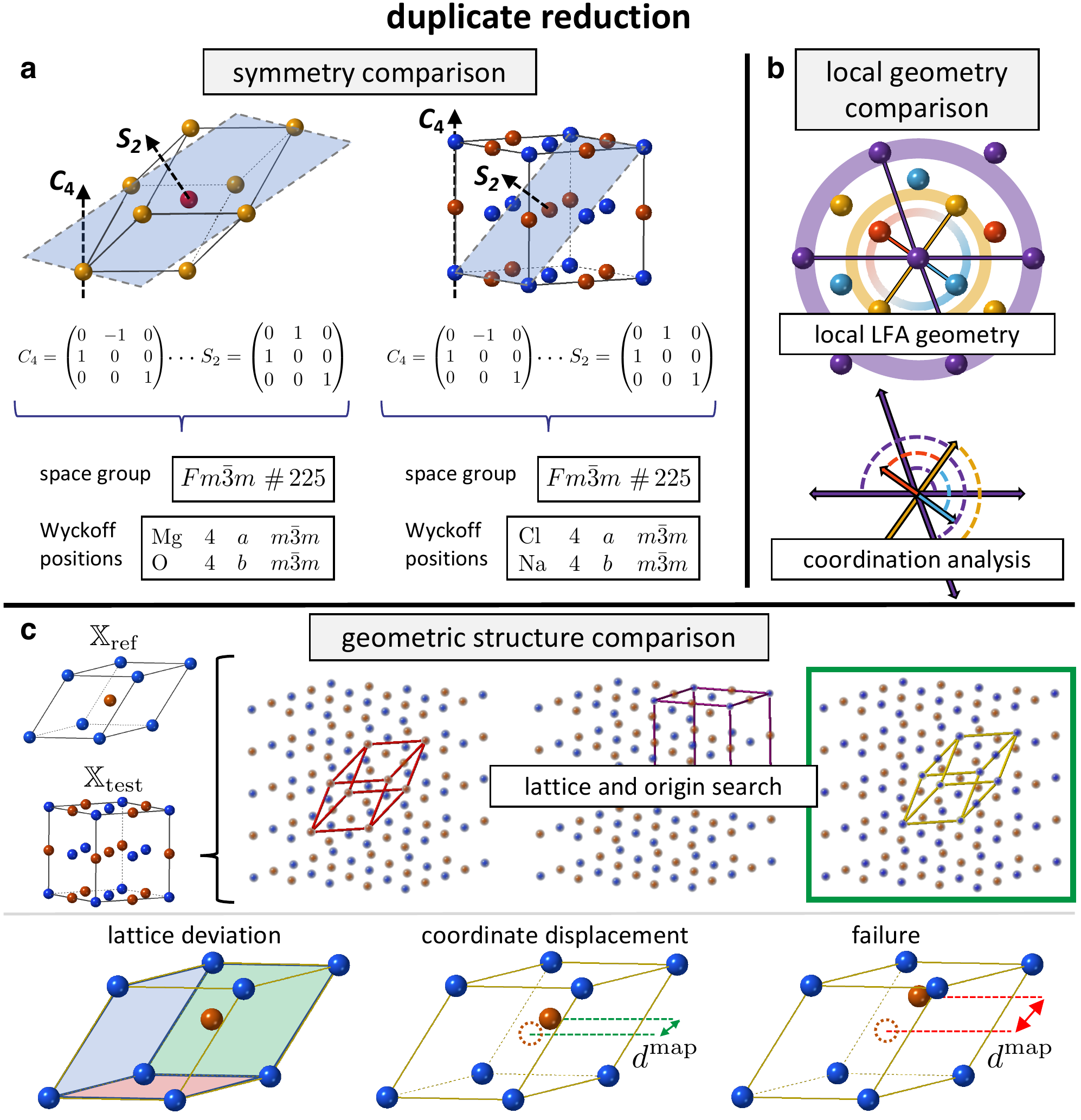}
    \caption{\textbf{Symmetry, local atomic geometry, and geometric structure comparisons in AFLOW-XtalFinder.}
    \small
    ({\textbf{a}}) Crystal isometries are calculated internally with AFLOW-SYM. 
    The space groups and occupied Wyckoff positions are compared, revealing isopointal structures.  
    ({\textbf{b}}) The local \underline{l}east-\underline{f}requently occurring \underline{a}tom (LFA) geometries 
    are computed and compared between structures. 
    An example local LFA geometry is shown for the quaternary Heusler structure~\cite{ABCD_cF16_216_c_d_b_a} (2-D projection),
    highlighting the closest neighbors (via solid lines) for each \LFA\ type to the central Mg atom (purple).
    Shaded concentric circles indicate the tolerance threshold for capturing atoms in the coordination shell
    with a thickness of 10\% of the distance from the central and connected atom.
    Local geometry vectors are compared against local geometries in other structures to determine mapping potential.
    ({\textbf{c}}) Two structures ($\mathbb{X}_{\mathrm{ref}}$ and $\mathbb{X}_{\mathrm{test}}$) 
    are mapped onto one another by 
    expanding $\mathbb{X}_{\mathrm{test}}$
    into a supercell and exploring commensurate lattice and origin choices with respect to $\mathbb{X}_{\mathrm{ref}}$.
    The yellow lattice (highlighted by the green box) is a potential match with $\mathbb{X}_{\mathrm{ref}}$.
    $\mathbb{X}_{\mathrm{test}}$ is transformed into the new representation 
    ($\widetilde{\mathbb{X}}_{\mathrm{test}}$), and
    the structures are quantitatively compared via the misfit criteria. 
    The structures are evaluated via their
    lattice deviation ($\epsilon_{\mathrm{latt}}$),
    coordinate displacement ($\epsilon_{\mathrm{coord}}$), and
    figure of failure ($\epsilon_{\mathrm{fail}}$).
    Distances between mapped atoms ($d^{\mathrm{map}}$) that are less than half the atom's nearest neighbor
    ($d_{\mathrm{nn}}/2$) are accounted for in the coordinate displacement (green dashed lines and arrows),
    while larger distances are described in the figure of failure (red dashed lines and arrows).
    }    
  \label{fig2_label}
\end{figure*}

{{
With a robust prototype classifier and generator in place,
comparison of prototypes is required to 
\textbf{i.} identify unique structure-types and 
\textbf{ii.} group similar ones together.
The prototype label and parameters alone cannot establish structural similarity
due to variations in the choice of lattice and origin, 
potentially affecting both the label ({\it{e.g.}} Wyckoff letters) and 
the parameters ({\it{e.g.}} lattice and non-fixed Wyckoff parameters).
Therefore, \AFLOWXTALFINDERSHORT\ offers three levels of comparison: symmetry, local atomic geometry, and 
complete crystal geometry.
They are described in the following three subsections. 
}

\label{subsec:symmetry}
\noindent\textbf{Isopointal structures: symmetry comparison}.
Symmetry analyses of crystals are required to 
identify structures of the same symmetry-type.
The isometries of crystals ({\it{e.g.}} rotations, roto-inversions, screw axes, and glide planes) are 
calculated via the routines of \AFLOWSYM~\cite{curtarolo:art135} to determine the space group and 
occupied Wyckoff positions (Figure~\ref{fig2_label}(a)). 
Results from \AFLOWSYM\ are robust against numerical tolerance issues and are consistent with experimentally determined 
symmetries in comparison to other symmetry software~\cite{curtarolo:art135}.

Crystals are isopointal if they
have commensurate space groups (equivalent or enantiomorphic pairs) and 
Wyckoff positions~\cite{Lima-De-Faria_StructureTypes_1990}.
Wyckoff positions are compatible if they have the same multiplicity and similar site symmetry designations.
Due to different setting and origin choices for the conventional cell, 
a strict site symmetry match is insufficient.
For instance, the Wyckoff positions with multiplicity 2 in space group \#47 ($Pmmm$) 
--- four $2mm$ (letters $i$-$l$), four $m2m$ (letters $m$-$p$), and four $mm2$ (letters $q$-$t$) --- 
form a Wyckoff set and are related via an automorphism of the space group operations~\cite{tables_crystallography,Boyle_WyckoffOrigin_1973,Koch_Automorphisms_1975}.
Depending on the assignment of the lattice parameters ($a$, $b$, and $c$) and origin choice, different 
--- and potentially equivalent --- Wyckoff decorations are possible.
Consequently, \AFLOWXTALFINDERSHORT\ tests permutations of the site symmetry symbol to expose 
positions that may be within the same Wyckoff set{ }~\footnote{Permuting the site symmetry symbol does not 
always reveal Wyckoff positions belonging to the same set since the site symmetry may originate from 
higher point symmetries (see example of space group \#66 ($Cccm$) and Wyckoff positions $i$ and $k$ in Ref.~\cite{Koch_Automorphisms_1975}).
Nevertheless, Wyckoff positions belonging to different sets cannot be matched, which will be revealed via the comparison.}. 

The symmetry calculation is performed automatically, 
{\it{i.e.}} it does not require input from the user.
Options are available to ignore symmetry and force geometric comparison of structures, 
which can identify crystals associated via symmetry subgroups. 

\label{subsec:environment_analysis}
\noindent\textbf{Isoconfigurational snapshots: local geometry comparison.} 
Beyond isopointal analyses, structures are further compared by inspecting arrangements of atoms, 
{\it{i.e.}} local atomic geometries.
Local geometry analyses have been fruitful in providing 
structural descriptors and similarity metrics between crystals of different types~\cite{curtarolo:art112,Zimmermann_LocalStructure_2019}.
However, the positions of these environments are often neglected, 
precluding the determination of one-to-one mappings between similar crystals.
Nevertheless, the analysis quickly identifies local geometries and is employed here 
to analyze structures beyond symmetry considerations
({\it{i.e.}} isoconfigurational {\it{versus}} isopointal~\cite{Lima-De-Faria_StructureTypes_1990}). 

Rather than determine the complete local atomic geometry for each atom, 
\AFLOWXTALFINDERSHORT\ builds a reduced representation:
neighborhoods comprised of only {{the \underline{l}east \underline{f}requently occurring \underline{a}tom (LFA) type(s)}}. 
The local LFA geometry analysis {{provides the connectivity for a subset of atoms ({\it{i.e.}} LFA-type) to discern 
if patterns}} are present in both structures, regardless of cell choice and crystal orientation. 
{{This description is preferred over the full local geometry}} because it is
\textbf{i.} computationally less expensive to calculate and 
\textbf{ii.} generally less sensitive to coordination cutoff tolerances. %(cutoff or nn calc) 
The latter is attributed to the fact that LFA geometries are more sparse. 

An example of a local LFA geometry is shown in Figure~\ref{fig2_label}(b).
A local LFA \underline{a}tomic \underline{g}eometry ($AG$) is 
{{a set of vectors connecting a central atom ($c$) to its closest neighbors}}:
\begin{equation}
    AG_{c}\equiv\{\mathbf{d}^{\mathrm{min}}_{ic}\} \, \, \forall i \, | \, {\mathrm{atom}}_{i} \in \{\mathrm{LFA(s)}\}, % n_{a}\mathbf{a} + n_{b}\mathbf{b} + n_{c}\mathbf{c} )||.
\end{equation}
where $\mathbf{d}^{\mathrm{min}}_{ic}$ {{is the minimum distance vector to the $i$ atom --- restricted to LFA-type(s) only --- % used to be ":"  
and is calculated via the method of images for periodic systems~\cite{Hloucha_minimumimage_1998}}}:
\begin{equation}
    %d^{\mathrm{min}}_{ic}=\min_{i}(||\min_{n_{a},n_{b},n_{c}}(\mathbf{x}_{i} - \mathbf{x}_{c} + n_{a}\mathbf{a} + n_{b}\mathbf{b} + n_{c}\mathbf{c} )||).
    d^{\mathrm{min}}_{ic}=\min_{i}(\min_{n_{a},n_{b},n_{c}}||(\mathbf{x}_{i} - \mathbf{x}_{c} + n_{a}\mathbf{a} + n_{b}\mathbf{b} + n_{c}\mathbf{c} )||).
\end{equation}
{{Here, $n_{a}$, $n_{b}$, and $n_{c}$ are the lattice dimensions along the lattice vectors $\mathbf{a}$, $\mathbf{b}$, and $\mathbf{c}$; 
and $\mathbf{x}_{i}$ and $\mathbf{x}_{c}$ are the Cartesian coordinates of the $i$ and $c$ (center) atoms, respectively.}}
A coordination shell with a thickness of $d^{\mathrm{min}}_{ic}/10$ captures other atoms of the same type 
to control numerical noise in the atomic coordinates (a similar tolerance metric is defined 
in \AFLOWSYM, {\it{i.e.}} loose preset tolerance value~\cite{curtarolo:art135}).  
This cutoff value yields expected coordination numbers for well-known systems and is 
comparable to results provided by other atom environment calculators~\cite{curtarolo:art112,Zimmermann_LocalStructure_2019}. 
If there is only one LFA type --- {\it{e.g.}} Si in $\alpha$-cristobalite (SiO$_{2}$)~\cite{A2B_tP12_92_b_a} ---
then the distance to the closest neighbor of that LFA type is calculated.
If there are multiple LFA types --- {\it{e.g.}} four for the quaternary Heusler (as illustrated in Figure~\ref{fig2_label}(b)) --- 
then the minimum distances to each LFA type are computed.
The local atomic geometry is calculated for each atom of the LFA type(s) in the unit cell, resulting in a list of 
atomic geometries ($\{AG_{c}\}$).
Therefore, $\alpha$-cristobalite has a set of four Si LFA geometries (one for each Si in the unit cell: 
{{\{$AG_{\mathrm{Si,1}}$, $AG_{\mathrm{Si,2}}$, $AG_{\mathrm{Si,3}}$, $AG_{\mathrm{Si,4}}$\}}}) and 
the quaternary Heusler has a set of four LFA geometries (one for each element type: 
{{\{$AG_{\mathrm{Au}}$, $AG_{\mathrm{Li}}$, $AG_{\mathrm{Mg}}$, $AG_{\mathrm{Sn}}$\}, respectively}}).

To investigate structural compatibility, local atomic geometry lists for compounds are compared.
In general, the local geometry comparisons err on the side of caution. 
For instance, comparing the cardinality of the coordination is often too strict.  
Despite a more sparse geometry space, slight deviations in position can move atoms
outside the coordination shell threshold, changing the atom cardinality and neglecting 
potential matches.
Local atomic geometries are thus compatible if
\textbf{i.} the central atoms are comparable types ({\it{i.e.}} same element and/or stoichiometric ratio in crystal), 
\textbf{ii.} the neighborhood of surrounding atoms have distances that match within 
20\% after normalizing with respect to $\max(AG_{c})$ ({\it{i.e.}} the largest distance in the local geometry cluster), and
\textbf{iii.} the angles formed by two atoms and the center atom match within 10 degrees.
To further alleviate the coordination problem, an exact geometry match is not required, 
{\it{i.e.}} some distances and angles are permitted to be missing.
Grouping local atomic geometries as compatible is favored 
to mitigate false negatives for equivalent structures.

% Could mention example: local atomic geometry filtered metal-carbides significantly

\label{subsec:lattice_origin_search}
\noindent\textbf{Isoconfigurational structures: Geometric structure comparison.} To resolve a commensurate representation between two structures for 
geometric comparison, 
one structure --- the reference $\mathbb{X}_\mathrm{ref}$ --- remains fixed and 
the other structure --- the potential duplicate $\mathbb{X}_\mathrm{test}$ --- is expanded into a supercell.  
Lattice vectors are identified within the supercell and compared against the reference structure.  
For any similar lattices to $\mathbb{X}_\mathrm{ref}$, 
$\mathbb{X}_\mathrm{test}$ is transformed into the new lattice representation ($\widetilde{\mathbb{X}}_\mathrm{test}$).
Origin shifts for this cell are then explored in an attempt to match atoms.
If one-to-one atom mappings exist between the two structures, then the similarity 
is quantified with the crystal misfit method (see ``Quantitative similarity measure'' subsection)~\cite{Burzlaff_ActaCrystA_1997}.
Misfit values below a given threshold indicate equivalent structures and the 
search terminates.
Alternatively, misfit values larger than the threshold are 
disregarded and the search continues until all lattices and origin shifts are exhausted.
The procedure is detailed below and an illustration of the process is depicted in Figure~\ref{fig2_label}(c).

The lattice search algorithm begins by scaling the volumes of the unit cells to 
compare structures with different volumes 
(an option is available to quantify the similarity between structures at fixed volumes).
Once scaled, the routine searches for translation vectors by generating a lattice grid of $\mathbb{X}_\mathrm{test}$.
The size of the grid is defined to encompass a sphere with a radius ($r_{\mathrm{grid}}$) 
equal to the maximum lattice vector length of $\mathbb{X}_\mathrm{ref}$, {\it{i.e.}}
\begin{equation}
  \label{eqn:grid_radius}
    r_{\mathrm{grid}} \equiv \max\left(a,b,c\right).
\end{equation}
Similar to a procedure described in Ref.~\cite{curtarolo:art135}, the necessary grid dimensions  
are given by the set of vectors perpendicular to each pair of $\mathbb{X}_\mathrm{ref}$ lattice vectors scaled by the grid radius
({\it{e.g.}} $\mathbf{n}_{1}=r_{\mathrm{grid}}\left(\mathbf{b}\times\mathbf{c}/||\mathbf{b}\times\mathbf{c}||\right)$). 
The scaled vectors are then transformed into the lattice basis ($\mathbf{L}$), 
via $\mathbf{n}'=\mathbf{L}^{-1}\mathbf{n}$, and the ceiling of the $\mathbf{n}'$ components indicate the grid dimensions:
$n_{a,b,c}=\mathrm{ceil}(\mathbf{n}_{a,b,c}')$.
The grid dimensions span between $-n_{a,b,c}\to n_{a,b,c}$
to account for different orientations/rotations between the structures.
To optimize the lattice search, translation vectors are explored in a grid comprised of only the 
LFA-type in $\mathbb{X}_\mathrm{test}$ 
(since they are the minimal set of atoms exhibiting crystal periodicity).
{{
In addition to verifying crystal periodicity,
candidate lattice vectors must be similar to those in the $\mathbb{X}_\mathrm{ref}$ lattice based on
\textbf{i.} lattice vector moduli ($\Delta l$),
\textbf{ii.} angles formed between pairs of lattice vectors ($\Delta \theta$) and
\textbf{iii.} volumes enclosed by three lattice vectors ($\Delta V$).
The tolerances values ($\Delta l$, $\Delta \theta$, $\Delta V$) 
are chosen based on how much the lattices are allowed to differ.
If the lattices are significantly different, then the lattice is ignored
(see the ``lattice deviation'' in the ``Quantitative similarity measure'' subsection).
Additionally, as a speed increase, commensurate lattices are sorted by minimum lattice deviation to find matches more quickly.
Upon finding a similar cell to $\mathbb{X}_\mathrm{ref}$, $\mathbb{X}_\mathrm{test}$ is transformed 
into the new lattice representation $\widetilde{\mathbb{X}}_\mathrm{test}$ and is stored if the 
representations have the same number of atoms (and types).}}

For each prospective unit cell, possible origin choices are explored.  
The origin of $\mathbb{X}_\mathrm{ref}$ is placed on one of the LFA-type atoms, and 
the origin of $\widetilde{\mathbb{X}}_\mathrm{test}$ is cycled through all atoms of its LFA-type.
Given an origin choice, a mapping procedure is attempted for all atoms in the unit cell.
The minimum Cartesian distance --- via the method of images for periodic systems~\cite{Hloucha_minimumimage_1998} ---
is determined for every atom $i$ in $\mathbb{X}_\mathrm{ref}$ to each atom $j$ in $\widetilde{\mathbb{X}}_\mathrm{test}$
\begin{equation}
  \label{eqn:distance_mappings}
    %d_{ij} = ||\min_{n_{a},n_{b},n_{c}}(\mathbf{x}_{i}-\mathbf{x}_{j} + n_{a}\mathbf{a} + n_{b}\mathbf{b} + n_{c}\mathbf{c} )||,
    d_{ij} = \min_{n_{a},n_{b},n_{c}}||(\mathbf{x}_{i}-\mathbf{x}_{j} + n_{a}\mathbf{a} + n_{b}\mathbf{b} + n_{c}\mathbf{c} )||,
\end{equation}
where $n_{a}$, $n_{b}$, and $n_{c}$ are the lattice dimensions along the lattice vectors $\mathbf{a}$, $\mathbf{b}$, and $\mathbf{c}$; 
and $\mathbf{x}_{i}$ and $\mathbf{x}_{j}$ are the Cartesian coordinates of the $i$ and $j$ atoms, respectively.
Given the set of distances $\{d_{ij}\}$, the minimum distance over all $j$ atoms is identified as the 
mapping distance, {\it{i.e.}}
\begin{equation}
  \label{eqn:mapped_distance}
  d^{\mathrm{map}}_i \equiv \min_{j}\{{d_{ij}\}}, 
\end{equation}    
regardless of the element type.
Once $d^{\mathrm{map}}_i$ is computed for all $i$, the following conditions are verified:
\textbf{i.} one-to-one mappings ({\it{i.e.}} no duplicate $j$ indices between $i$ indices), and
\textbf{ii.} no cross-matching between element types ({\it{i.e.}} cannot map a single element type to multiple types in $\widetilde{\mathbb{X}}_\mathrm{test}$).
If either condition is violated, the mappings are ignored and the search continues.

Given a successful mapping, the similarity of the two crystals in the corresponding representations 
are quantified, indicating equivalent or unique structures. 
If no mapping is found for any lattice and origin choice, then the structures are considered distinct
and are not assigned a similarity value.

\label{sec:misfit_criteria}
\noindent\textbf{Quantitative similarity measure.} To compare two crystals in a given representation, a method proposed by \MISFITAUTHORS\ is employed~\cite{Burzlaff_ActaCrystA_1997}.
The similarity between structures is quantified by a misfit value, $\epsilon$, which incorporates differences between 
lattice vectors and atomic coordinates via~\cite{Burzlaff_ActaCrystA_1997}:
\begin{equation}
  \label{eqn:misfit}
    \epsilon \equiv 1.0 - \left(1.0 - \epsilon_{\mathrm{latt}}\right) \left(1.0 - \epsilon_{\mathrm{coord}}\right) \left(1.0 -\epsilon_{\mathrm{fail}}\right).
\end{equation}
The misfit quantity is bound between zero and one: 
structures with a value close to zero match and 
those with a value close to one do not match.
Special misfit ranges defined by \MISFITAUTHORS\ are adopted here~\cite{Burzlaff_ActaCrystA_1997}
\begin{align}
  \label{eqn:misfit_ranges}
  0 < \epsilon \leq {{\epsilon_{\mathrm{match}}}} &\colon \mbox{match}, \nonumber \\
  {{\epsilon_{\mathrm{match}}}} < \epsilon \leq {{\epsilon_{\mathrm{family}}}} &\colon \mbox{same family, and} \\
  {{\epsilon_{\mathrm{family}}}} < \epsilon \leq 1 &\colon \mbox{no match} \nonumber. 
\end{align}
The ``same family'' designation generally corresponds to crystals with common symmetry subgroups.
{{\MISFITAUTHORS\ recommend $\epsilon_{\mathrm{match}}=0.1$ and $\epsilon_{\mathrm{family}}=0.2$
based on definitions from Pearson~\cite{Pearson_ChemPhysMetals_1972} and Parth{\'e}~\cite{Parthe_StructuralChemistry_1990}.
In the ``Finding $\epsilon_{\mathrm{match}}$: structural misfit versus calculated property ($\Delta{}H_{\mathrm{atom}}$)'' section,
heuristic misfit thresholds are identified based on the allowed maximum enthalpy differences between similar structures.}} 

The deviation of the lattices, $\epsilon_{\mathrm{latt}}$, captures the difference between the lattice face diagonals of 
$\widetilde{\mathbb{X}}_\mathrm{test}$ and $\mathbb{X}_\mathrm{ref}$~\cite{Burzlaff_ActaCrystA_1997}
\begin{align}
  \label{eqn:lattice_deviation}
  \epsilon_{\mathrm{latt}} &\equiv 1 - (1 - D_{12})(1 - D_{23})(1 - D_{31}), \\ 
    D_{kl} &\equiv \frac{  ||\widetilde{\mathbf{d}}_{kl}^{\mathrm{test}} - \mathbf{d}_{kl}^{\mathrm{ref}}||
                    + ||\widetilde{\mathbf{f}}_{kl}^{\mathrm{test}} - \mathbf{f}_{kl}^{\mathrm{ref}}|| }
                 {||\mathbf{{d}}_{kl}^{\mathrm{ref}} - \mathbf{f}_{kl}^{\mathrm{ref}}||},
\end{align}
where $\mathbf{f}_{kl}$ and $\mathbf{d}_{kl}$ denote the diagonals by adding and subtracting, respectively, 
the $k$ and $l$ lattice vectors.
{{In the lattice search algorithm, $\Delta l$, $\Delta \theta$, and $\Delta V$ tolerances 
are coupled to $\epsilon_{\mathrm{latt}}$, and are tuned to ensure $\epsilon_{\mathrm{latt}}\le\epsilon_{\mathrm{family}}$.
}} 

The coordinate deviation --- measuring the disparity between atomic positions in the two structures --- 
is based on the mapped atom distances 
($d^{\mathrm{map}}_{i}$ or $d^{\mathrm{map}}_{j}$ as computed with Equations~(\ref{eqn:distance_mappings}) and (\ref{eqn:mapped_distance})) 
and the atoms' nearest neighbor distances in the respective structures, $d_{\mathrm{nn}}$~\cite{Burzlaff_ActaCrystA_1997}
\begin{align}
  \label{eqn:coordinate_displacement}
    \epsilon_{\mathrm{coord}} &\equiv \frac{\sum_{i}^{{\widetilde{N}}^{\mathrm{test}}} \left(1 - \widetilde{n}_{i}^{\mathrm{test}}\right) d_{i}^{\mathrm{map}}
                    + \sum_{j}^{{N}^{\mathrm{ref}}} \left(1 - n_{j}^{\mathrm{ref}}\right) d_{j}^{\mathrm{map}}}
                   {\sum_{i}^{{\widetilde{N}}^{\mathrm{test}}} \left(1 - \widetilde{n}_{i}^{\mathrm{test}}\right) d_{\mathrm{nn},i}^{\mathrm{test}}
                    + \sum_{j}^{{N}^{\mathrm{ref}}} \left(1 - n_{j}^{\mathrm{ref}}\right) d_{\mathrm{nn},j}^{\mathrm{ref}}}.
\end{align}
$\widetilde{N}^{\mathrm{test}}$ and $N^{\mathrm{ref}}$ are the number of atoms in the two crystals.
If $d^{\mathrm{map}} < d_{\mathrm{nn}}/2$, then a ``switch'' variable $n$ is set to zero and the  
mapped atom distance is included in $\epsilon_{\mathrm{coord}}$.
Otherwise, $n$ is set to one, signifying the mapped atoms are far apart and not considered in $\epsilon_{\mathrm{coord}}$.
These atoms are represented in the figure of failure, $\epsilon_{\mathrm{fail}}$~\cite{Burzlaff_ActaCrystA_1997} 
\begin{align}
  \label{eqn:figure_of_failure}
    \epsilon_{\mathrm{fail}} &\equiv \frac{\sum_{i}^{\widetilde{N}^{\mathrm{test}}} \widetilde{n}_{i}^{\mathrm{test}} +\sum_{j}^{N^{\mathrm{ref}}} n_{j}^{\mathrm{ref}}}
                              {\widetilde{N}^{\mathrm{test}} + N^{\mathrm{ref}}}.
\end{align}

{{
Other metrics can be used to assess structural similarity,
including the \underline{r}oot \underline{m}ean \underline{s}quare ($rms$) of the atom 
positions~\cite{pymatgen-structure-matcher_2011} and
coordination characterization functions~\cite{spap_jpcm_2017}.
\AFLOWXTALFINDERSHORT\ employs the crystal misfit criteria to
incorporate structural differences between both the lattice and atom positions.
Differences between common similarity metrics ---
and their software implementations ---
are discussed in more detail in the ``Comparison Accuracy'' subsection.
}}

\begin{figure*}[]
  \centering
  \includegraphics[width=1\textwidth]{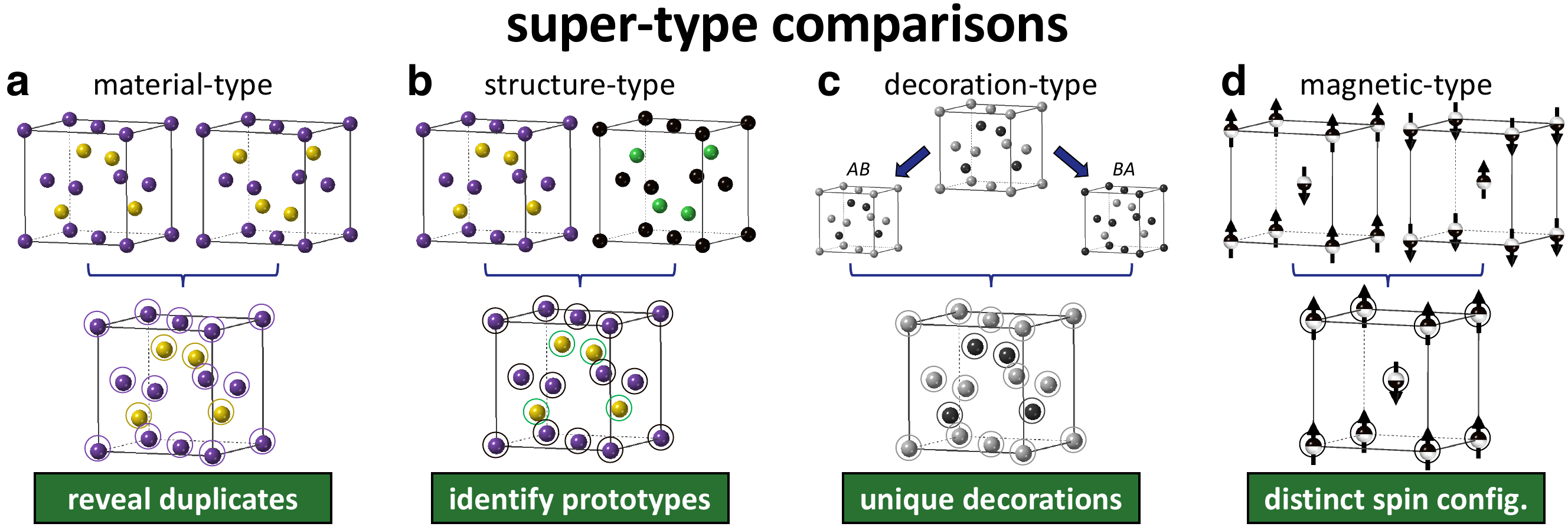}
    \caption{\textbf{Available super-type comparison modes.}
    \small
    (\textbf{a}) Material-type: maps same element types, revealing duplicate compounds.
    (\textbf{b}) Structure-type: maps structures regardless of the element types, identifying crystallographic prototypes.
    (\textbf{c}) Decoration-type: creates and compares all atom decorations for a given structure, determining unique and equivalent decorations (in this case, atom decorations $AB$ and $BA$ match).
    (\textbf{d}) Magnetic-type: maps compounds by element types and magnetic moments, discerning distinct spin configurations.
    }
  \label{fig3_label}
\end{figure*}

\noindent\textbf{Super-type comparisons.} To explore new areas of materials space, the \AFLOWXTALFINDERSHORT\ module
\textbf{i.} identifies equivalent and unique materials, 
\textbf{ii.} uncovers common structure-types across different compounds ({\it{i.e.}} prototypes),
\textbf{iii.} determines inequivalent atom decorations for a given crystal structure, and
\textbf{iv.} discerns distinct magnetic structure configurations.
The corresponding comparison modes are denoted as
{material-,} {structure-,} {decoration-,} and {magnetic-type,} respectively (Figure~\ref{fig3_label}). % {} for new-line wrapping
Each variant uses the underlying procedures discussed in the ``Results'' section 
({\it{i.e.}} symmetry, local atomic geometry, and geometric structure comparisons) with
different restrictions on mapping atom types.

\noindent\textbf{Material-type.} Material-type comparisons map atoms of the same atomic species (Figure~\ref{fig3_label}(a)).
For example, given two ZnS zincblende compounds~\cite{AB_cF8_216_c_a}, a material-type comparison maps Zn$\to$Zn and S$\to$S in the 
two structures.
Therefore, the method reveals duplicate compounds within a data set.

\noindent\textbf{Structure-type.} Conversely, structure-type comparisons ignore atomic species and map any atom-type 
with compatible stoichiometric ratios (Figure~\ref{fig3_label}(b)).
In the case of zincblende structures ZnS and SiC, a structure-type comparison attempts to map Zn$\to$Si and S$\to$C, or 
Zn$\to$C and S$\to$Si, since the compounds are equicompositional.
This mode exposes unique backbone structures and is practical for crystallographic prototyping.
Identifying prototypes is also useful for modeling solid solutions and disordered materials~\cite{curtarolo:art151,Oses_NatureReview_2020}.  

\noindent\textbf{Decoration-type.}
\label{subsubsec:decoration_type}
The decoration-type (or permutation-type) mode determines 
unique atom decorations for a given crystal structure, 
{\it{i.e.}} inequivalent colorings of a structure, where each element is denoted by a different color (Figure~\ref{fig3_label}(c)). 
Continuing with the zincblende example, the $A$ and $B$ atomic sites are equivalent:
swapping elements on the sites results in a duplicate compound  
compared to the original decoration. %, resulting in degenerate properties.
Thus, only one site decoration choice is necessary to create a distinct compound.
Given a compound with $n$ species, there are $n!$ possible atom permutations.
\AFLOWXTALFINDERSHORT\ automatically 
\textbf{i.} generates compounds with the different atom decorations for a crystal, 
\textbf{ii.} compares the decorations (via a material-type comparison), and 
\textbf{iii.} identifies the unique configurations. 
Atom decorations are only compared if atomic types have the same 
Wyckoff multiplicity and similar site symmetries (see subsection ``Isopointal structures: symmetry analysis'').

\noindent Equivalent decoration groups need to obey Lagrange's theorem~\cite{Beachy_AbstractAlgebra_2006}: 
the order $h$ of subgroup $H$ divides the group $G$ with order $g$ ({\it{i.e.}} $\mathrm{mod}(g,h)=0$). 
Accordingly, the numbers of unique and equivalent decorations must divide the total number of decorations, 
{\it{i.e.}} satisfy divisor theory.
The possible equivalent decoration groups --- out of $n!$ -- are dictated by its divisors, and are 
enumerated below for $2<n<5$ (elemental compounds, $n=1$, are excluded):
\begin{myitemize_nobullet}
  \item{2! = 2  : 2, 1}
  \item{3! = 6  : 6, 3, 2, 1}
  \item{4! = 24 : 24, 12, 8, 6, 4, 3, 2, 1}
  \item{5! = 120 : 120, 60, 40, 30, 24, 20, 15, 12, 10, 8, 6, 5, 4, 3, 2, 1}
\end{myitemize_nobullet}
For example, the possible groupings for a ternary compound ($n=3$) are:
6, 3, 2, and 1 unique sets with 1, 2, 3, and 6 decorations per set, respectively.

\begin{figure*}[]
  \centering
  \includegraphics[width=1\textwidth]{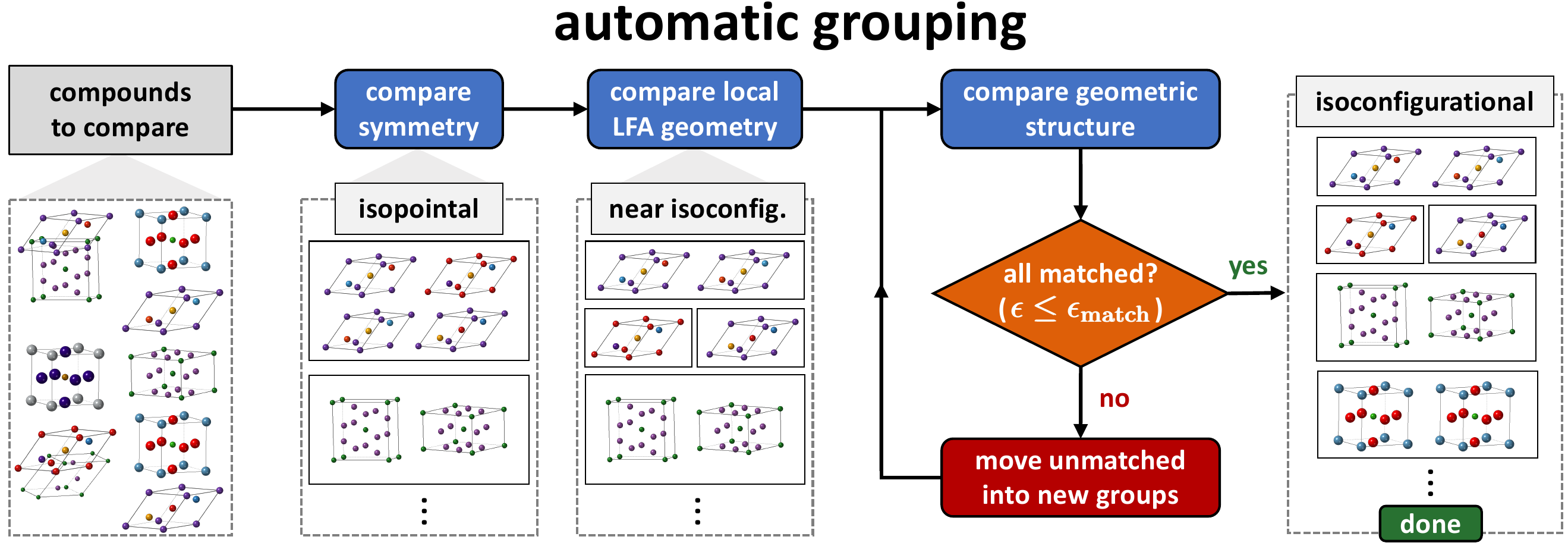}
    \caption{\textbf{Automatic grouping of multiple compounds.}
    \small
    {{
    Compounds are compared in the following sequence: symmetry, local LFA geometry, and geometric structure.
    The algorithms determine isopointal, near isoconfigurational, and isoconfigurational structures, respectively, and
    aggregate them into similar sets (enclosed in black solid-lined boxes).
    Unmatched structures ({\it{i.e.}} $\epsilon>\epsilon_{\mathrm{match}}$) after the initial geometric structure comparison are put into new groups and re-compared until
    all equivalent structures are grouped.
    This sequence is the same for material-, structure-, decoration-, and magnetic-type comparisons;
    however, the criteria for atom mappings differ (see subsection ``Super-type comparisons'' for details).
    }}
    The symmetry, local LFA geometry, and geometric structure comparisons
    {{(blue boxes)}} are multithreaded for parallel computation.
    }
  \label{fig4_label}
\end{figure*}

\noindent Depending on the matching (misfit) tolerance and the choice of the reference decoration, 
calculated equivalency groups can violate divisor theory.
For instance, two decorations can match with a certain misfit; 
however, a better match with a smaller misfit can exist with another decoration. 
To combat incorrect groupings, \AFLOWXTALFINDERSHORT\ executes a consistency check, verifying the groupings 
are commensurate with the possible divisors.
If they are not, \AFLOWXTALFINDERSHORT\ searches for better matches and regroups the compatible decorations.

\noindent For example, \ICSD\ entry BiITe \#10500 (original geometry) has six possible atom decorations: 
$ABC$, $BAC$, $CBA$, $ACB$, $CAB$, and $BCA$.
Since the three equicompositional sites are comprised of the same Wyckoff multiplicity and site symmetry 
(multiplicity 1 and site symmetry 3m. in space group \#156), 
all structures are placed in the same initial comparison group, 
with $ABC$ chosen as the reference decoration (since it is the first in the set).
After comparing, the equivalent groups and their misfit values are:
\begin{itemize}
    \item{$ABC$ = $BAC$ ($\epsilon=0.0889$) = $CBA$ ($\epsilon=0.0144$),}
    \item{$CAB$ = $ACB$ ($\epsilon=0.0889$), and}
    \item{$BCA$ (no equivalent decorations).}
\end{itemize}
However, the number of equivalent decorations in each set are not the same, 
violating Lagrange's theorem~\cite{Beachy_AbstractAlgebra_2006}.
Furthermore, all misfits values should be the same, since the underlying structure is unchanged.
The incommensurate groupings are a symptom of only comparing to the reference decoration, 
as opposed to cross-comparing with other decorations.

\noindent To remedy incorrect groupings, \AFLOWXTALFINDERSHORT\ checks for better matches 
({\it{i.e.}} potential equivalent decorations with lower misfit values).
Therefore, the ``duplicate'' decorations are compared to the other reference decorations 
and regrouped to minimize the misfit value.
In this case, the subsequent cross-comparisons are performed:
\begin{itemize}
    \item{$BAC$ with $CAB$ and $BCA$,}
    \item{$CBA$ with $CAB$ and $BCA$, and}
    \item{$ACB$ with $BCA$ (not compared with $ABC$; performed previously).}
\end{itemize}
Consequently, the final equivalent decorations are 
\begin{itemize}
    \item{$ABC$ = $CBA$ ($\epsilon=0.0144$),}
    \item{$CAB$ = $BAC$ ($\epsilon=0.0144$), and}
    \item{$BCA$ = $ACB$ ($\epsilon=0.0144$).}
\end{itemize}
The groupings above satisfy Lagrange's theorem, and the equivalent structures in each group have the same misfit value with respect to their reference decoration.

\noindent\textbf{Magnetic-type.}\label{subsubsec:magnetic_type} 
Magnetic-type comparisons map atoms of the same atomic species and similar magnetic moments, 
{\it{i.e.}} analyzes spin configurations (Figure~\ref{fig3_label}(d)).
For instance, given two body-centered cubic chromium compounds with antiferromagnetic ordering, 
the routine attempts to map Cr$^{\uparrow}$$\to$Cr$^{\uparrow}$ and 
Cr$^{\downarrow}$$\to$Cr$^{\downarrow}$.
A magnetic moment tolerance threshold denotes equivalent spin sites; 
where the default tolerance is $0.1\mu_\mathrm{B}$. 
The analysis can be performed for both collinear and non-collinear systems.
The magnetic-type comparison can be joined with a magnetic structure generator 
to create distinct spin configurations for high-throughput simulation.

\label{subsec:high-throughput_comparisons}
\noindent\textbf{Multiple comparisons.} With the plethora of compounds generated by computational frameworks 
--- such as \AFLOW~\cite{aflow_fleet_chapter_full,curtarolo:art142}, 
\NOMAD~\cite{nomadMRS}, 
Materials Project~\cite{APL_Mater_Jain2013},
High-Throughput Toolkit~\cite{httk},
Materials Cloud/\AIIDA~\cite{Pizzi_AiiDA_2016},
and \OQMD~\cite{Saal_JOM_2013} ---
{{automatically}} comparing structures is necessary for high-throughput classification of unique/duplicate compounds and structure-types.
For this purpose, we developed an automatic comparison procedure for multiple crystals (Figure~\ref{fig4_label}).
{{
Compounds are first grouped into isopointal sets by analyzing and comparing the symmetries of the structures, 
aggregating them by stoichiometry, space groups, and Wyckoff sets (calculated via \AFLOWSYM~\cite{curtarolo:art135}).
Next, compounds are further partitioned into near isoconfigurational sets by determining and comparing the 
local \LFA\ geometries in each structure. 
Within each near isoconfigurational group}}, one representative structure --- generally the first in the set --- 
is compared to the other structures via geometric comparisons and the misfit values are stored.
Once the comparisons finish, any unmatched structures 
({\it{i.e.}} misfit values greater than {{$\epsilon_{\mathrm{match}}$}}) are reorganized 
into new comparison sets.
The process repeats until all structures have been assembled into matching groups or all comparison pairs are exhausted.
{{
The three comparison analyses are performed in this order for two reasons:
\textbf{i.} to categorize structural similarity to varying degrees 
(isopointal, near isoconfigurational, and isoconfigurational) and
\textbf{ii.} to efficiently group compounds to reduce the computational cost of the geometric structure comparison 
(see ``Speed and scaling considerations'' in the Discussion). 
This procedure is the same for material-, structure-, decoration-, and magnetic-type comparisons; however, different 
atom mapping restrictions are applied depending on the comparison mode.
}}

\noindent{{\textbf{Multithreading.} 
To enhance calculation speed, multithreading capabilities can be employed.  
The three computationally intensive procedures 
--- calculating the symmetry, constructing the local \LFA\ geometry, and performing geometric comparisons --- 
are partitioned onto allocated threads, offering significant speed increases for large collections of structures.
}}

\noindent\textbf{Automatic comparisons.}
There are three built-in functions to compare multiple structures automatically: 
\textbf{i.} compare structures provided by a user,
\textbf{ii.} compare an input structure to prototypes in \AFLOW~\citeANRL, and
\textbf{iii.} compare an input structure to entries in the \AFLOW.org repository.
An overview of each high-throughput method is discussed below and usage is detailed in the Methods section.

\noindent\textbf{Compare user datasets.}
Users can load crystal geometries and compare them automatically with \AFLOWXTALFINDERSHORT.
Options to perform both material-type and structure-type comparisons are available to identify unique/duplicate compounds or prototypes, respectively.
For structure-type comparisons, the unique atom decorations for each representative structure are determined.
Once the analysis is complete, \AFLOWXTALFINDERSHORT\ groups compatible structures together and returns the corresponding misfit values.

% compare2prototypes
\noindent\textbf{Compare to \AFLOW\ prototypes libraries.}
Given an input structure, this routine returns similar \AFLOW\ prototype(s) along with their misfit value(s) (Figure~\ref{fig5_label}(a)).
\AFLOW\ contains structural prototypes that can be rapidly decorated for high-throughput materials discovery: 
590 in the \PROTOTYPEENCYCLOPEDIA~\citeANRL{} and  
1,492 in the High-throughput Quantum Computing library~\cite{curtarolo:art65}.
In this method, \AFLOW\ prototypes are extracted --- based on similar stoichiometry, space group, and Wyckoff positions to the input --- and compared to the user's structure. 
Since only matches to the input are relevant, the procedure terminates before regrouping any unmatched prototypes. 
The attributes of matched prototypes are also returned, including the prototype label, 
mineral name, {\it{Strukturbericht}} designation, and links to the corresponding \PROTOTYPEENCYCLOPEDIA\ webpage.
The scheme identifies common structure-types with the \AFLOW\ libraries or --- if no matches are found --- reveals new prototypes.
Absent prototypes can be characterized automatically in the \AFLOW\ standard designation with \AFLOWXTALFINDERSHORT{}'s 
prototyping tool (discussed in subsection ``\IDEALPROTOTYPESECTION'').

\begin{figure*}[]
  \centering
  \includegraphics[width=1\textwidth]{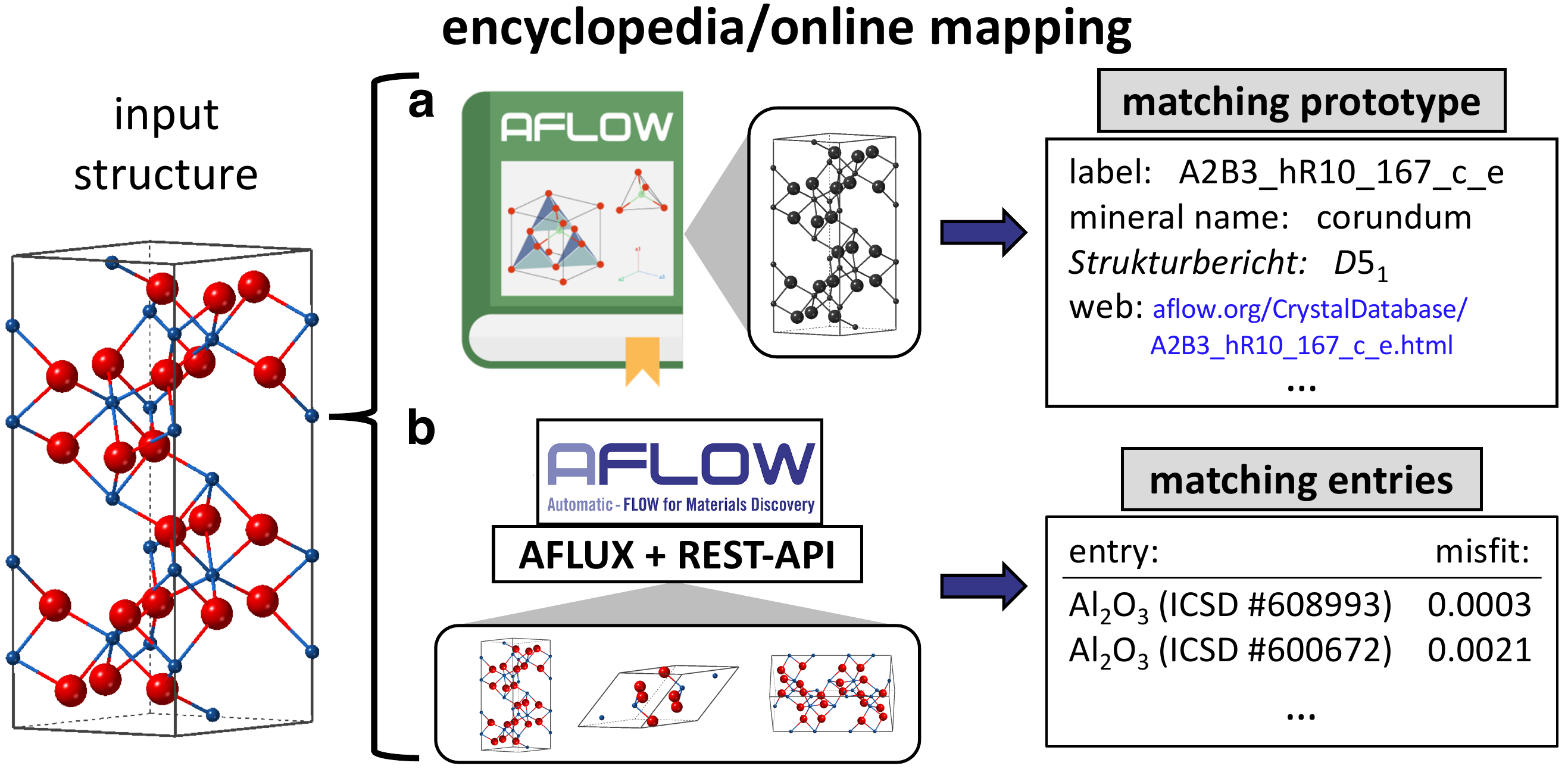}
    \caption{\textbf{Encyclopedia/online prototype mapping.}
    \small
    An input Al$_{2}$O$_{3}$ (corundum) compound is compared to entries in
    (\textbf{a}) \AFLOW\ \PROTOTYPEENCYCLOPEDIA\ and
    (\textbf{b}) the \AFLOW.org repository.
    Potential equivalent entries are retrieved automatically from the respective catalog and compared with \AFLOWXTALFINDERSHORT.
    Matching entries and their level of similarity (misfit) are returned.
    }
  \label{fig5_label}
\end{figure*}

\begin{table}
    \centering
    \caption{\textbf{AFLOW.org entries equivalent to an input sodium chloride (rocksalt) compound.}
    \small
      A list of equivalent compounds to the \PROTOTYPEENCYCLOPEDIA's rocksalt structure with the default degrees of freedom 
      (label=AB\_cF8\_225\_a\_b, parameters=5.64~{\AA}).  
      The compound name, auid, misfit ($\epsilon$), and enthalpy per atom ($H_{\mathrm{atom}}$) are listed for all similar structures in the database. 
      Volume scaling is suppressed for the comparison to incorporate volume differences.
      The first 25 and last 2 entries are from \AFLOW's \ICSD\ and \LIBTWO\ catalogs, respectively.
    }
    \begin{tabular}{l|l|r|r}
        \multicolumn{1}{c|}{\multirow{ 2}{*}{compound}} & \multicolumn{1}{c|}{\multirow{ 2}{*}{auid}} & \multicolumn{1}{c|}{\multirow{ 2}{*}{$\epsilon$}} & \multirow{ 2}{*}{\shortstack[c]{$H_{\mathrm{atom}}$ \\ (eV/atom)}} \\
        & & & \\
      \hline
        ClNa   &   aflow:d241535faf2a4519   &   0.00514317   &  -3.39101  \\
        ClNa   &   aflow:82a178672a734c47   &   0.00537828   &  -3.39082  \\
        ClNa   &   aflow:1cd71114972d46dd   &   0.00514250   &  -3.39078  \\
        ClNa   &   aflow:d0c93a9396dc599e   &   0.00496982   &  -3.39075  \\
        ClNa   &   aflow:5699b196418c6044   &   0.00450831   &  -3.39066  \\
        ClNa   &   aflow:9017f9c64ead22ab   &   0.00434390   &  -3.39062  \\
        ClNa   &   aflow:39ab5e62afdb5ac0   &   0.00429571   &  -3.39062  \\
        ClNa   &   aflow:c16c0f1c061f7d3e   &   0.00424606   &  -3.39061  \\
        ClNa   &   aflow:2f4b5e32510830a0   &   0.00427698   &  -3.39061  \\
        ClNa   &   aflow:b5ab343f3a484538   &   0.00421818   &  -3.39060  \\
        ClNa   &   aflow:cc41860d69de2888   &   0.00405508   &  -3.39056  \\
        ClNa   &   aflow:b2ec4b68e12f3674   &   0.00404585   &  -3.39056  \\
        ClNa   &   aflow:4f19021768a3118a   &   0.00399452   &  -3.39055  \\
        ClNa   &   aflow:ec23029a18d3fec9   &   0.00373730   &  -3.39049  \\
        ClNa   &   aflow:a4652bde28e67c3d   &   0.00339698   &  -3.39041  \\
        ClNa   &   aflow:18ebb85b07a92f89   &   0.00386555   &  -3.39033  \\
        ClNa   &   aflow:1354bbef4edd80b3   &   0.00383458   &  -3.39032  \\
        ClNa   &   aflow:182f848dd10cc403   &   0.00383093   &  -3.39031  \\
        ClNa   &   aflow:d996b8d524516c24   &   0.00380678   &  -3.39030  \\
        ClNa   &   aflow:a5755554aaf5d10e   &   0.00379748   &  -3.39030  \\
        ClNa   &   aflow:9466351a9cbac2c9   &   0.00379835   &  -3.39030  \\
        ClNa   &   aflow:9a28207fd647e477   &   0.00379460   &  -3.39029  \\
        ClNa   &   aflow:e3e31c4914d59e25   &   0.00379517   &  -3.39029  \\
        ClNa   &   aflow:fd711a60dbfba2de   &   0.00378193   &  -3.39028  \\
        ClNa   &   aflow:55d2cbd0f4018884   &   0.00405470   &  -3.39013  \\
        ClNa   &   aflow:3bd528dd9f88be7d   &   0.00395044   &  -3.39233  \\
        ClNa   &   aflow:f4b806d73482566c   &   0.00345690   &  -3.39121  \\
    \end{tabular}
    \label{table:database_comparison}
\end{table}

% compare2database 
\noindent\textbf{Compare to \AFLOW.org repository.}
Compounds are compared to entries in the \AFLOW.org repository using the \AFLOW\ \REST- and \AFLUX\ Search-\APIS~\cite{curtarolo:art92,curtarolo:art128} (Figure~\ref{fig5_label}(b)).
An \AFLUX\ query ({\it{i.e.}} matchbook and directives) is generated internally 
and returns database compounds similar to the input structure based on species, stoichiometry, space group, and Wyckoff positions.
With the \AURL\ from the \AFLUX\ response, structures for the entry are retrieved via the \RESTAPI.
The most relaxed structure is extracted by default; however, options are available to obtain structures at different  
{\it{ab-initio}} relaxation steps.
The set of entries from the database are then compared to the input structure.
Similar to the \AFLOW\ prototype comparisons, candidate entries are only compared against the input structure,
{\it{i.e.}} the procedure terminates without regrouping unmatched entries.

With the underlying \AFLUX\ functionality, material properties can also be extracted, 
highlighting the structure-property relationship amongst similar materials.
For instance, the enthalpy per atom ($H_{\mathrm{atom}}$) for matching database entries are printed by including the 
\verbWrap|enthalpy_atom| \API\ keyword in the query.
Any number or combination of properties can be queried; available \API\ keywords are located in  
Refs.~\cite{curtarolo:art92,curtarolo:art128}.
Table~\ref{table:database_comparison} shows the comparison results between a rocksalt NaCl compound and matching DFT-relaxed structures in the \AFLOW.org repository along 
with their misfits and enthalpies per atom.

This routine reveals equivalent \AFLOW.org compounds, if similar materials exists in                                              
the database.  
As such, it can estimate structural properties {\it{a priori}}; before performing any calculations.
The estimation is based on the following assumptions: 
\textbf{i.} the matching \AFLOW\ material resides at a local minimum in the energy landscape and 
\textbf{ii.} the input structure relaxes to the same geometry as that \AFLOW\ compound, given comparable 
calculation parameters.
The functionality can explore properties that are not calculated for a given entry, but are calculated for an equivalent entry. 
For example, compounds in \AFLOW's prototype catalogs (\LIBONE, \LIBTWO, \LIBTHREE, etc.) do not usually have band structure data; however, 
corresponding \ICSD\ entries can be found which do provide band structure information.
Finally, the method can identify compounds that are absent from the database and prioritize them for future calculation, 
enhancing the diversity of the \AFLOW.org repositories.

% moved prototyping section further up (Prototyping functionality)

\begin{figure*}[]
  \centering
  \includegraphics[width=1\textwidth]{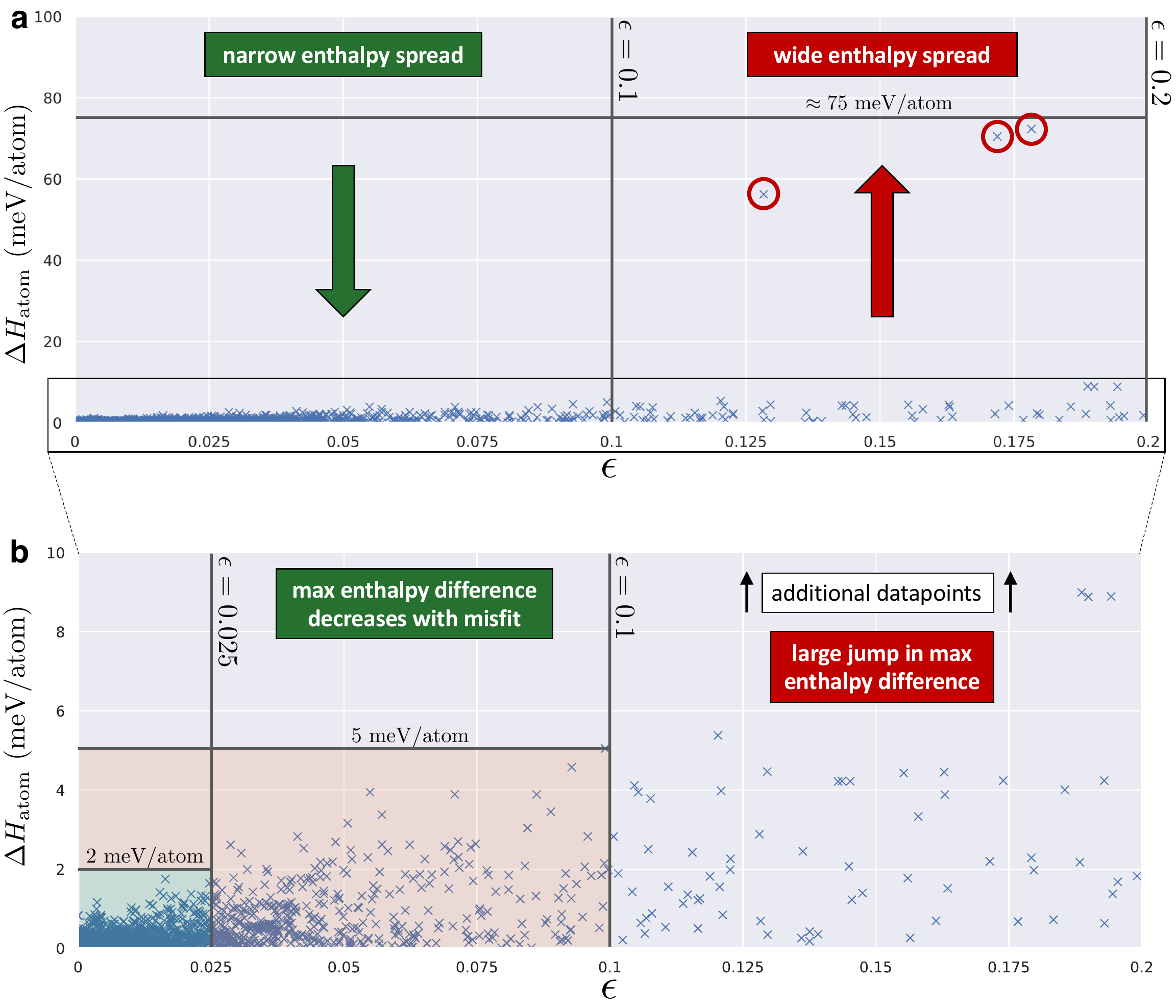}
    \caption{\textbf{Enthalpy difference per atom and misfit value between compared structures in the \AFLOW-\ICSD\ catalog.}
    \small
    The misfit value ($\epsilon$) and the difference in enthalpy per atom 
    ($\Delta{}H_{\mathrm{atom}}=H_{\mathrm{atom}}^{\mathrm{ref}}-H_{\mathrm{atom}}^{\mathrm{test}}$) 
    {{for all \AFLOW-\ICSD\ entries with similar parameters are shown above.  
    The plots show the misfit values between $0-0.2$ with two enthalpy difference ranges (for clarity): 
    (\textbf{a}) $0-100$~meV/atom and 
    (\textbf{b}) $0-10$~meV/atom.
    The plot with the full misfit domain and enthalpy range is shown in the \SUPPLEMENTARYMATERIALS.
    Candidate misfit thresholds are chosen based on the acceptable maximum enthalpy deviation between match structures.  
    For example, misfit values below $\epsilon=0.025$ and $\epsilon=0.1$ (black vertical lines) 
    are expected to yield enthalpy differences no larger than  
    $\Delta{}H_{\mathrm{atom}}\approx2$~meV/atom and
    $\Delta{}H_{\mathrm{atom}}\approx5$~meV/atom, respectively (black horizonal lines).
    As the misfit values increases beyond $\epsilon>0.1$,
    the spread of the data points also increases.
    A large jump in the maximum enthalpy difference occurs at approximately $\epsilon=0.125$, indicating matched structures
    near this value and beyond are not guaranteed to have similar enthalpies.
    }}
    }
  \label{fig6_label}
\end{figure*}

\label{sec:usage}
\label{sec:using_aflowxtalmatch}                           
\noindent\textbf{Using \AFLOWXTALFINDER.} 
For ease-of-use, the \AFLOWXTALFINDERSHORT\ routines are accessible via 
a command-line interface and a Python environment (see Methods for details).

\noindent{{\textbf{Ideal prototype analysis in \AFLOW.org.}}}
{{
The ideal prototype designations 
--- for both the original and relaxed geometries --- 
have been successfully determined for all 4+ million entries in the \AFLOW.org repository.
The prototype label, parameter variables, and parameter values are incorporated 
into the \AFLOW\ \REST- and Search-\API{}s~\cite{curtarolo:art92,curtarolo:art128}.
The corresponding \API\ keywords for the original geometries are}}
\begin{myitemize}
\item{\verbWrapBlue!aflow_prototype_label_orig!,}
\item{\verbWrapBlue!aflow_prototype_params_list_orig!, {{and}}}
\item{\verbWrapBlue!aflow_prototype_params_values_orig!.}
\end{myitemize}
{{For the DFT-relaxed geometries, the keywords are}}  
\begin{myitemize}
\item{\verbWrapBlue!aflow_prototype_label_relax!,}
\item{\verbWrapBlue!aflow_prototype_params_list_relax!, {{and}}}
\item{\verbWrapBlue!aflow_prototype_params_values_relax!.}
\end{myitemize}

{{
The prototype keywords enable researchers to search for materials by structure.
This new feature is useful for 
identifying possible crystal structures given experimental data.
For example, with composition, space group, and occupied Wyckoff information
(characteristics often known to experimentalists); 
users can construct the corresponding prototype label(s) and extract 
all compounds based on the provided structure-type. 
The keywords are also used to identify the frequency of certain prototypes
in the \AFLOW.org repository.
For example, all compounds that are isopointal to the corundum prototype 
(labels: A2B3\_hR10\_167\_c\_e and A3B2\_hR10\_167\_e\_c) can be retrieved for 
both the original and relaxed geometries.
Moreover, this search capability is used to discern if a structure-type is novel or 
has been reported previously.
}}

{{
The ideal prototype keywords also reveal
whether a compound retains the same prototype designation before and after relaxation.
For structures that retain the same prototype label, the 
parameter values show the continuous structure transition during relaxation.
For structures that transform into different prototypes, the symmetry-based designations 
highlight the symmetries that were broken. % upon relaxation.
This can indicate that certain element combinations/arrangements
are averse to certain prototype structures.
More advanced relaxation techniques, 
{\it{e.g.}} symmetry-constrained relaxations~\cite{Lenz_SymConstrain_2019},
would be required to restrict the relaxation to a given prototype structure.
}}

\noindent\textbf{{{Finding $\epsilon_{\mathrm{match}}$: structural misfit versus calculated property ($\Delta{}H_{\mathrm{atom}}$).}}}
{{To identify a suitable threshold for matching similar structures ($\epsilon_{\mathrm{match}}$),}}
Figure~\ref{fig6_label} plots the misfit value 
($\epsilon$) between two mapped structures and their difference in enthalpy per atom ($\Delta{}H_{\mathrm{atom}}$).
The structures in the test set are comprised of DFT-relaxed entries from the 
{{entire \AFLOW-\ICSD\ catalog as of \AFLOWICSDDATEXTALMATCH\ (60,390)~\cite{ICSD,curtarolo:art75}}}. 
{{Compounds are grouped via commensurate atomic elements, stoichiometries, symmetries, and local \LFA\ geometries.}}
{{Furthermore, only}} compounds calculated with similar {\it{ab initio}} settings are compared together 
--- such as \LDAU\ parameters, \underline{kp}oint \underline{p}er \underline{r}eciprocal \underline{a}tom (KPPRA), and pseudopotentials 
(see \SUPPLEMENTARYMATERIALS\ for details) ---
to prevent extraneous enthalpy differences due to differing parameters.
In addition, magnetic systems are excluded since the magnetic moment is not {{incorporated into
the misfit value}}. 
For these comparisons, the unit cell volumes are not rescaled, and
the best lattice/origin choices are explored (minimizing the misfit value)
to show better correlation with the enthalpies.
{{After grouping the structures and identifying one-to-one mappings, misfit values for the remaining \NUMBEROFENTHALPYCOMPARISONS\ comparison pairs are calculated.
Figures~\ref{fig6_label}(a) and (b) show the enthalpy difference ranges $0-100$~meV/atom and $0-10$~meV/atom, respectively, highlighting the maximum enthalpy differences at different misfit values.
}}

\setlength{\tabcolsep}{4pt}
\begin{table*}
    \caption{\textbf{Functionalities of comparison codes specific to high-throughput analysis and structure prototyping.}
     \small
    This tabulation is not exhaustive; many programs offer additional analyses, such as 
    fragment/molecular comparisons, and are outside the scope of this work.
    $^{\dagger}$: optional symmetry input.
    $^{*}$: requires symmetry input.
    $^{\S}$: \STRUCTUREMATCHER\ compares to the \AFLOW\ \PROTOTYPEENCYCLOPEDIA\ partially, as it does not provide the internal degrees of freedom for the prototype.
    $^{\|}$: \STRUCTUREMATCHER\ matches magnetic structures with opposite spins (\texttt{SpinComparator} function).
    }
    \resizebox{\linewidth}{!}{%
    \begin{tabular}{l c c c c c c c c}
        & \multicolumn{1}{c}{\multirow{ 3}{*}{\shortstack[c]{AFLOW- \\ XtalFinder}}} & \multicolumn{1}{c}{\multirow{ 3}{*}{\shortstack[c]{Structure \\ Matcher}}} & \multicolumn{1}{c}{\multirow{ 3}{*}{\shortstack[c]{XTALCOMP}}} & \multicolumn{1}{c}{\multirow{3}{*}{SPAP}} & \multicolumn{1}{c}{\multirow{ 3}{*}{{CMPZ}}} & \multicolumn{1}{c}{\multirow{ 3}{*}{{CRYCOM}}} & \multicolumn{1}{c}{\multirow{ 3}{*}{\shortstack[c]{STRUCTURE- \\ TIDY}}} & \multicolumn{1}{c}{\multirow{ 3}{*}{{COMPSTRU}}}  \\
        \\
        \\
        \hline                            % AFLOW       % STRUCTUREMATCHER      % XTALCOMP        %SPAP              % CMPZ               % CRYCOM           %STRUCTURE-TIDY      % COMPSTRU
        source-code available          & \AFLOWXMARK\ & \AFLOWXMARK\            & \AFLOWXMARK\ &              &                         &                   &                   &                   \\
        prototyping tools              & \AFLOWXMARK\ &                         &              &              &                         &                   &                   &                   \\
        consider symmetry              & \AFLOWXMARK\ &                         &              &              & \AFLOWXMARK$^{\dagger}$ & \AFLOWXMARK$^{*}$ & \AFLOWXMARK$^{*}$ & \AFLOWXMARK$^{*}$ \\
        perform multiple comparisons   & \AFLOWXMARK\ & \AFLOWXMARK\            &              &              &                         &                   &                   &                   \\
        material-type comparisons      & \AFLOWXMARK\ & \AFLOWXMARK\            & \AFLOWXMARK\ & \AFLOWXMARK\ & \AFLOWXMARK\            & \AFLOWXMARK\      & \AFLOWXMARK\      & \AFLOWXMARK\      \\
        structure-type comparisons     & \AFLOWXMARK\ & \AFLOWXMARK\            & \AFLOWXMARK\ & \AFLOWXMARK\ & \AFLOWXMARK\            & \AFLOWXMARK\      & \AFLOWXMARK       & \AFLOWXMARK\      \\
        decoration-type comparisons    & \AFLOWXMARK\ &                         &              &              &                         &                   &                   &                   \\
        magnetic-type comparisons      & \AFLOWXMARK\ & \AFLOWXMARK$^{\|}$  &              &              &                         &                   &                   &                   \\
        compare to database            & \AFLOWXMARK\ &                         &              &              &                         &                   &                   &                   \\
        compare to prototypes          & \AFLOWXMARK\ & \AFLOWXMARK$^{\S}$      &              &              &                         &                   &                   &                   \\
        quantitative similarity metric & \AFLOWXMARK\ & \AFLOWXMARK\            &              & \AFLOWXMARK\ & \AFLOWXMARK\            & \AFLOWXMARK\      & \AFLOWXMARK\      & \AFLOWXMARK\      \\
    \end{tabular}
    }
    \label{table:code_comparisons}
\end{table*}
\setlength{\tabcolsep}{2pt}

{{
In general, the misfit value correlates with the enthalpy difference for $\epsilon\le0.1$:
as the misfit value decreases, the enthalpy difference also reduces.
For $\epsilon>0.1$, the enthalpy spread widens with the misfit.
Some comparison-pairs exhibit large misfit values, but have similar enthalpies.
This follows intuition since it is possible for significantly differing structures to have 
similar properties.
The sparsity of the data points for large values of $\epsilon$ is attributed to the lack of 
one-to-one mappings as structures become increasingly dissimilar.
This suggests \AFLOWXTALFINDERSHORT\ and the misfit criteria are better suited to 
quantifying similar structures, rather than relating disparate structures.  
}}

{{{
Figure~\ref{fig6_label} reveals possible thresholds for $\epsilon_{\mathrm{match}}$
based on the maximum enthalpy difference allowed for mapped structures.
For $\epsilon\le0.1$, the enthalpy differences per atom are all below 5~meV/atom, with the 
exception of one comparison-pair. 
Reducing the misfit cutoff reasonably guarantees the enthalpy differences will also decrease; 
{\it{e.g.}} enthalpies will be within 1~meV/atom and 2~meV/atom for misfit values below $3.58\times10^{-3}$ and 0.025, respectively.
The maximum enthalpy difference jumps significantly (to approximately 50~meV/atom) near $\epsilon=0.125$.
Thus, matching structures with misfits beyond $\epsilon=0.1$ are not guaranteed to exhibit similar enthalpies.
This value is in agreement with \MISFITAUTHORSPOSSESIVE\ proposed threshold. 
}}
{{
By default, \AFLOWXTALFINDERSHORT\ employs a threshold of $\epsilon_{\mathrm{match}}=0.1$ 
to ensure similar materials match within approximately 5~meV/atom.
The threshold is also used for comparing prototypes; two 
prototypes that match within $\epsilon_{\mathrm{match}}=0.1$, 
are expected to have enthalpies within 5~meV/atom when decorated with the same atomic elements.
Users can adjust to stricter (or looser) thresholds for matching;
however, $\epsilon_{\mathrm{match}}\gg0.1$ is not guaranteed to yield small enthalpy differences between
matched structures.
}}

\noindent\textbf{Functionality differences with other codes.} 
In addition to \AFLOWXTALFINDERSHORT, other structure comparison tools are available 
to the materials science community:
\STRUCTUREMATCHER~\cite{pymatgen-structure-matcher_2011},
\XTALCOMP~\cite{xtal-comp_2012},
\SPAP~\cite{spap_jpcm_2017},
\CMPZ~\cite{Hundt_CMPZ_2006},
\CRYCOM~\cite{crycom_1994},
\STRUCTURETIDY\ (via \PLATON)~\cite{structure-tidy_1987}, and
\COMPSTRU~\cite{Flor_COMPSTRU_2016}.
A summary of the offered functionalities related to automatic comparisons
and structure prototyping
is indicated in Table~\ref{table:code_comparisons} and described below.

% Discuss availability of source code
\noindent\textbf{Source code availability.} The source codes are available for the following packages:
\AFLOWXTALFINDERSHORT\ (via \AFLOW), 
\STRUCTUREMATCHER\ (via Pymatgen), and
\XTALCOMP\ (via \XTALOPT).
Pre-compiled binaries for \SPAP\ on different operating systems are available with the \CALYPSO\ software~\cite{Wang_StructurePrediction_PRB_2010,Wang_CALYPSO_CPC_2012}.
Source codes for the other software: 
\CMPZ\ (implemented in \KPLOT~\cite{Hundt_KPLOT_1979}), 
\CRYCOM, 
\STRUCTURETIDY, and 
\COMPSTRU\ (online) are not available.
Therefore, the latter packages are not convenient for merging into user-workflows.

% Discuss input formats
\noindent\textbf{Input file formats.} 
{{
The different structure file formats for each comparison code are listed below;
\AFLOWXTALFINDERSHORT: 
\VASP\ (\POSCAR)~\cite{kresse_vasp}, 
\FHIAIMS~\cite{Blum_CPC2009_AIM}, 
\QUANTUMESPRESSO~\cite{quantum_espresso_2009}, 
\ABINIT~\cite{gonze:abinit}, 
\ELK~\cite{Elk_LAPW}, and
\CIF;
\STRUCTUREMATCHER: \POSCAR, \CIF, \ABINIT, and a Pymatgen object;
\XTALCOMP: C++ object (and \POSCAR{}s via online tool);
\SPAP: \POSCAR\ (and \CIF{}s via \CIFTWOCELL~\cite{Bjorkman_cif2cell_2011});
\CMPZ: \KPLOT\ structure files;
\CRYCOM: \FDAT\ files (native to the \underline{C}ambridge \underline{S}tructural \underline{D}atabase (\CSD)~\cite{Groom_CSD_2016});
\STRUCTURETIDY: creates structures based on space group input, unit cell parameters, and positions of atoms; and
\COMPSTRU: \CIF.
}}

% Discuss symmetry differences
\noindent\textbf{Symmetry analysis.} \AFLOWXTALFINDERSHORT\ is the only package coupled with an internal symmetry 
calculator (\AFLOWSYM~\cite{curtarolo:art135}).
\CRYCOM, \STRUCTURETIDY, and \COMPSTRU\ require a symmetry input (space group number) to perform 
the comparison, but lack methods to calculate the symmetry internally.
\CMPZ\ allows a symmetry input; however it is not required to perform the comparison.
\STRUCTUREMATCHER, \XTALCOMP, and \SPAP\ do not consider symmetry in their structural analyses.

% Discuss performing multiple comparisons
\noindent\textbf{Multiple comparisons.} The only packages that offer comparison of multiple materials in a single command 
are \AFLOWXTALFINDERSHORT\ and \STRUCTUREMATCHER.  
Other software, such as \XTALCOMP\ and \SPAP, showcase comparison results performed on multiple structures, 
but multi-comparison routines are not available to users. 
To achieve similar functionality, users need to implement external regrouping procedures.

% Discuss decoration-type comparisons
\noindent\textbf{Decoration-type comparisons.} \AFLOWXTALFINDERSHORT\ is the only code that 
automatically determines the unique (and equivalent) atom decorations for a given crystal structure.
With other packages, users must externally generate, organize, and compare the subsequent decorations.
Beyond the lack of routines to generate decorations, the codes find incorrect equivalent decoration groups.  
In the BiITe (ICSD \#10500) example discussed in the ``Decoration-type'' comparison mode section,    
\STRUCTUREMATCHER's \verbWrap!group_structures! function (with \verbWrap!ltol!=0.2, \verbWrap!stol!=0.17, \verbWrap!angle_tol!=5.0) 
identifies groupings that violate divisor theory: 
\begin{itemize}
\item{$ABC$ = $BAC$ ($rms$=0.1196) = $CBA$ ($rms$=0.0194),}
\item{$CAB$ = $ACB$ ($rms$=0.1196), and}
\item{$BCA$ (no equivalent decorations).}
\end{itemize}
A similar discrepancy occurs for the remaining codes depending on the 
structure input order and comparison tolerance(s). 
\AFLOWXTALFINDERSHORT\ checks consistency with Lagrange's theorem to validate permissible 
decoration groupings; a burden that falls to users of the other packages. 

Furthermore, \AFLOWXTALFINDERSHORT\ calculates Wyckoff positions {\it{a priori}} to check if decorations 
are commensurate based on symmetry, {\it{i.e.}} mapped positions have the same multiplicity and similar site symmetries. 
Without this validation, positions with differing site symmetries can be mistaken as equivalent. 
For example, the GaPu compound (\ICSD\ \#103930) has three Wyckoff positions: 
Ga 8 $h$ $m.2m$, 
Pu 4 $d$ $-4m2$, and 
Pu 4 $e$ $4mm$ (before and after relaxation).  
From symmetry, the Ga and Pu sites cannot be swapped to yield degenerate compounds.  
Despite the symmetry restrictions, \STRUCTUREMATCHER\ incorrectly groups the decorations (\verbWrap!group_structures!) into equivalent bins using their 
default tolerance values ({\it{i.e.}} \verbWrap!ltol!=0.2, \verbWrap!stol!=0.3, \verbWrap!angle_tol!=5). 
\AFLOWXTALFINDERSHORT\ --- with its symmetry analysis coupled with mapping routines --- 
correctly distinguishes these decorations and is the only viable option to establish consistent unique/duplicate decorations for crystalline prototypes.

% Discuss comparisons to database
\noindent\textbf{Database comparisons.} 
\AFLOWXTALFINDERSHORT\ is the only module that features a method for comparing input structures to 
a database of materials, namely \AFLOW.org.  
The \API\ functionality coupled with 
\AFLOWXTALFINDERSHORT\ ensures comparisons are performed with the most current 
version of the database, incorporating new materials as they are calculated. 
Furthermore, \AFLOWXTALFINDERSHORT\ users can compare structures at various relaxation steps 
(with the \verbWrap!--relaxation_step! option).
Users of the other packages need to extract 
relevant structures ({\it{e.g.}} similar compositions, space group, Wyckoff positions, 
and local atomic geometries at particular relaxation steps) 
by-hand or code auxiliary scripts to perform similar functionality.

% Discuss comparison to prototypes: 
\noindent\textbf{Prototype comparisons.} 
\AFLOWXTALFINDERSHORT\ compares to all prototype structures in \AFLOW: 
the \PROTOTYPEENCYCLOPEDIA, the High-Throughput Quantum Computing library, and initial geometries in the 
\AFLOW.org repository.
Similar to the database comparisons, \AFLOWXTALFINDERSHORT\ automatically includes 
new prototype structures as they are added to \AFLOW.   
\STRUCTUREMATCHER\ only compares structures against a static subset of \AFLOW\ prototypes 
({\it{i.e.}} the \PROTOTYPEENCYCLOPEDIA\ via \texttt{AflowPrototypeMatcher}~\cite{pymatgen_aflowprototypematcher}).
Moreover, \AFLOWXTALFINDERSHORT\ provides the internal degrees of freedom for any structure
(via the prototyping routines); functionality all existing codes currently lack. 
To compare to these prototype representations, users of other packages need to convert the degrees of freedom 
-- including expansion of the corresponding Wyckoff positions --- 
into a structure file {\it{a priori}}.

% Discuss speed
\noindent\textbf{Speed and scaling considerations.} Comparison speeds were evaluated for packages that could be 
compiled locally on a Linux machine: 
\AFLOWXTALFINDERSHORT\ (V\AFLOWVERSION), 
\STRUCTUREMATCHER\ (V\STRUCTUREMATCHERVERSION), and 
\XTALCOMP\ (\XTALCOMPVERSION).  
The benchmarks were run with a single processor on a 2.60~GHz Intel(R) Xeon(R) Gold 6142 CPU machine, 
and the respective default tolerances were used for all codes.  
Pairwise comparison times are similar between the packages; on the order of milliseconds. 
For a 1,494 pairwise comparison test 
set{ }
~\footnote{The test set is comprised of ICSD unaries (original geometries) with 1-105 
atoms per unit cell and varying symmetries; along with a mix of equivalent and inequivalent structures.}, 
\AFLOWXTALFINDERSHORT\ averaged 282 milliseconds/comparison, 
\STRUCTUREMATCHER\ averaged 689 milliseconds/comparison, and 
\XTALCOMP\ averaged 33 milliseconds/comparison.  
\XTALCOMP\ is the fastest, but at the cost of limited scope and functionality: \XTALCOMP\ does not scale volumes and quits immediately if the 
volumes/lattice vectors are different sizes.
Therefore, \XTALCOMP\ finds fewer matches (18), while \AFLOWXTALFINDERSHORT\ and \STRUCTUREMATCHER\ find more (approximately 450).  
For large, skewed cells, \AFLOWXTALFINDERSHORT\ can be slower since it does not convert to 
Minkowski, Niggli, and/or primitive cells
by default to preserve the input representations (unlike \STRUCTUREMATCHER\ and \XTALCOMP).  
To increase speed in \AFLOWXTALFINDERSHORT, lattice transformations are available with the relevant options for 
Minkowski (\verbWrap!--minkowski!), 
Niggli (\verbWrap!--niggli!), and 
primitive (\verbWrap!--primitive!) reductions.

For multiple comparisons, \AFLOWXTALFINDERSHORT\ scales more efficiently with the number of compounds 
when compared to other software. 
\AFLOWXTALFINDERSHORT\ groups structures into near isoconfigurational sets via symmetry and local atomic geometries 
(both calculated internally), eliminating unnecessary mapping comparisons between dissimilar structures. 
All other codes do not use symmetry or local geometry analyses to optimize groupings. 
\STRUCTUREMATCHER\ --- and other straightforward extensions to pairwise comparisons --- 
only groups by composition and executes more mapping procedures.  
For an ensemble of 600 structures modeling a disordered 5-metal carbide~\cite{curtarolo:art110,curtarolo:art140}, 
\AFLOWXTALFINDERSHORT\ partitions immediately into 54 groups via the symmetry and local geometry comparisons, 
while \STRUCTUREMATCHER\ puts all 600 structures into one large group. 
Consequently, \AFLOWXTALFINDERSHORT\ executes 546 structure mappings, and 
\STRUCTUREMATCHER\ performs 17,640 mapping attempts before arriving at the same solution.

While all benchmarks were performed serially, \AFLOWXTALFINDERSHORT\ routines are parallelized and users can 
specify the number of threads for the analyses (\verbWrap!--np!=$x$), offering additional speed over other packages
for large-scale {{automatic}} comparisons requiring little or no user input.  
Therefore, \AFLOWXTALFINDERSHORT\ will be more performant, especially when comparing more structures.

\noindent{\textbf{Comparison accuracy.} As shown in Figure~\ref{fig6_label}, 
the \AFLOWXTALFINDERSHORT\ misfit value decreases with the enthalpy difference between matched 
compounds, validating its accuracy.  
Comparisons with \STRUCTUREMATCHER\ are less accurate --- and at times qualitatively incorrect --- 
due to conversions of structures to an ``average lattice''~\cite{pymatgen-structure-matcher_2011}, 
matching significantly differing lattices with no penalty on the $rms$ value.
For example, Se (\ICSD\ \#104187, space group \#229) and Se (\ICSD\ \#57181, space group \#166) 
are classified as distinct by \AFLOWXTALFINDERSHORT\ because the lattices are considerably dissimilar 
($\epsilon_{\mathrm{latt}}=0.15$), consistent with the space groups.  
Despite having different symmetries, \STRUCTUREMATCHER\ inaccurately finds $rms=0$ between the structures.  
This distorts their $rms$ value, and it cannot be used to correlate properties of matched compounds, {\it{e.g.}} enthalpy.  
Conversely, \XTALCOMP\ is qualitatively accurate, but it lacks a quantitative similarity metric (the return type is a Boolean).  
Furthermore, \XTALCOMP\ comparisons neglect volume scaling between structures, an essential feature for identifying prototypes. 
\AFLOWXTALFINDERSHORT\ is the only comparison software suitable for quantitatively measuring similarity of materials and prototypes.

Overall, \AFLOWXTALFINDERSHORT\ is optimized for {{prototype detection and structural comparison}} within large datasets.
In addition, it is designed to be accessible to the broader materials science community for integration 
into user workflows.

%\subsection*{Prototypes in the AFLOW-ICSD catalog}
% orig  : entries=60,390 -> unique=34,820 (57.7%) -> prototypes=15,205 (25.2%) -> isopointal=9,060 (15.0%) (did not check automorphisms of Wyckoff positions)
% relax : entries=60,390 -> unique=33,544 (55.5%) -> prototypes=14,692 (24.3%) -> isopointal=9,032 (15.0%) (did not check automorphisms of Wyckoff positions)

% orig  : entries=60,390 -> unique=34,820 (57.7%) -> prototypes=15,205 (25.2%) -> isopointal=8,521 (14.1%) (checked automorphisms of Wyckoff positions)
% relax : entries=60,390 -> unique=33,544 (55.5%) -> prototypes=14,692 (24.3%) -> isopointal=8,493 (14.1%) (checked automorphisms of Wyckoff positions)

%\noindent\textbf{Prototypes in the \AFLOW-\ICSD\ catalog.} 
\noindent\textbf{{{Unique}} prototypes in the \AFLOW-\ICSD\ catalog.} 
With \AFLOWXTALFINDERSHORT, unique compounds and prototypes have been identified in the \ICSD\ catalog 
of the \AFLOW.org repository.
Table~\ref{table:ICSD_duplicates_and_prototypes} shows the statistics for 
the original (reported by the \ICSD~\cite{ICSD}) 
and DFT-relaxed geometries (via the \AFLOW\ standard~\cite{curtarolo:art104})
for 60,390 entries.
Material-type comparisons and suppressing volume scaling reveal the number of unique compounds.
Subsequent structure-type comparisons (allows for volume scaling) determine the number of distinct prototypes.
The representative compound for each prototype is chosen as the entry with the lowest \ICSD\ number, 
since it is generally the oldest among the compounds (and less likely to be removed from the \ICSD).
The unique atom decorations for each prototype are determined via the decoration-type comparison.
Moreover, the prototypes are cast into the \AFLOW\ prototype designation form, exposing its degrees of freedom.
Finally, the prototypes are compared to the \PROTOTYPEENCYCLOPEDIA~\citeANRL{} to 
distinguish between existing and new structures.
{{For the subsequent comparisons, the matching threshold is chosen as $\epsilon_{\mathrm{match}}=0.1$
to group similar compounds (and prototypes when decorated with alike atoms) that are expected to have enthalpies
differing by approximately 5~meV/atom or less (see subsection ``Finding $\epsilon_{\mathrm{match}}$: structural misfit versus calculated property ($\Delta{}H_{\mathrm{atom}}$)'' for details).}}

The analysis shows that the original geometry set includes 34,820 unique compounds (57.7\% of the total number of entries) and 15,205 prototypes (25.2\%).
Similarly, the DFT-relaxed set contains 33,544 unique compounds (55.5\%) and 14,692 prototypes (24.3\%).
Based on the symmetry comparisons, there are 8,521 (14.1\%) original and 8,493 (14.1\%) relaxed distinct isopointal structure-types. 
In general, the original geometry set has more distinct compounds and prototypes than the DFT-relaxed set.
This is attributed to the different volumes ({\it{e.g.}} measured temperatures and pressures) of the original geometries, 
while the DFT-relaxed geometries represent the ground state configurations, yielding additional degenerate compounds.

Overall, the binaries and ternaries have the highest number of entries, and thus, prototypes.
The number of entries/prototypes drops with species $n>3$, following statistics regarding the complexity of materials~\cite{Mackay_Complexity_2001}.
Table~\ref{table:ICSD_prototypes_by_symmetry} partitions the prototypes by their symmetry, {\it{i.e.}} Bravais lattices.
The number of lower symmetry prototypes (tri, mcl, and mclc) exceed the higher ones (cub, fcc, bcc) 
because lower symmetry classes have additional degrees of freedom, permitting more geometric diversity.
Similar to Table~\ref{table:ICSD_duplicates_and_prototypes}, there are generally more original prototypes 
in each lattice type compared to their relaxed counterparts.  
However, 347 structures changed lattice symmetry upon relaxation, yielding the following net Bravais lattice type gains/losses: 
tri (+8), 
mcl (-17), 
mclc (+46), 
orc (-31), 
orcc (+48), 
orcf (+1), 
orci (+5), 
tet (+17), 
bct (+8), 
hex (-87), 
rhl (-7), 
cub (+2), 
fcc (+4), 
bcc (+3).  
In particular, the mclc and orcc Bravais lattices had a considerable influx of prototypes, offsetting the expected 
reduction of prototypes due to DFT geometry optimization.

% Table separated via number of species
\begin{table}
  \caption{\textbf{Number of unique materials and prototypes in the AFLOW-ICSD repository.}
  \small
  The statistics are organized by number of 
  species, and the counts are shown for the original and DFT-relaxed entries.
  }
  \begin{tabularx}{\linewidth}{Y Y Y Y Y Y}
    \multicolumn{1}{c}{\multirow{2}[3]{*}{species}} & \multicolumn{1}{c}{\multirow{2}[3]{*}{entries}} & \multicolumn{2}{c}{unique materials} & \multicolumn{2}{c}{prototypes} \\
    \cmidrule(l){3-4} \cmidrule(l){5-6}
          &        & \multicolumn{1}{c}{original} & \multicolumn{1}{c}{relaxed} & \multicolumn{1}{c}{original} & \multicolumn{1}{c}{relaxed} \\
    \midrule
    1     &  1,606 &    538 &      440 &    236 &     196 \\
    2     & 22,530 &  9,050 &    8,569 &  3,140 &   3,017 \\
    3     & 26,109 & 17,285 &   16,725 &  6,419 &   6,168 \\ 
    4     &  8,185 &   6218 &    6,101 &  3,962 &   3,894 \\
    5     &  1,644 &   1442 &    1,426 &  1,177 &   1,156 \\
    6     &    291 &    262 &      258 &    246 &     236 \\ 
    7     &     25 &     25 &       25 &     25 &      25 \\ 
    \midrule
    total & 60,390 & 34,820 &   33,544 & 15,205 &  14,692   
  \end{tabularx}
  \label{table:ICSD_duplicates_and_prototypes}
\end{table}

% Table separated via symmetry
\begin{table}
  \caption{\textbf{The prototypes and their symmetries.}
  \small
  The prototypes are grouped into the 14  
  Bravais lattices: triclinic (tri), monoclinic, (mcl), base-centered monoclinic (mclc), orthorhombic (orc), 
  base-centered orthorhombic (orcc), face-centered orthorhombic (orcf), body-centered orthorhombic (orci), 
  tetragonal (tet), body-centered tetragonal (bct), hexagonal (hex), rhombohedral (rhl), simple cubic (cub), 
  face-centered cubic (fcc), and body-centered cubic (bcc).
  }
  \begin{tabularx}{\linewidth}{Y Y Y}
    \multicolumn{1}{c}{\multirow{2}[3]{*}{lattice type}} & \multicolumn{2}{c}{prototypes} \\
    \cmidrule(l){2-3}
    &        \multicolumn{1}{c}{original} & \multicolumn{1}{c}{relaxed} \\
    \midrule
    tri     & 1,345 & 1,338 \\ 
    mcl     & 2,266 & 2,255 \\ 
    mclc    & 2,165 & 2,195 \\ 
    orc     & 2,665 & 2,493 \\ 
    orcc    & 1,093 & 1,158 \\ 
    orcf    &   167 &   157 \\ 
    orci    &   305 &   292 \\ 
    tet     &   938 &   887 \\ 
    bct     &   861 &   817 \\ 
    hex     & 1,720 & 1,540 \\ 
    rhl     &   996 &   911 \\ 
    cub     &   274 &   265 \\ 
    fcc     &   227 &   211 \\ 
    bcc     &   183 &   173 \\ 
  \end{tabularx}
  \label{table:ICSD_prototypes_by_symmetry}
\end{table}

% Table of common prototypes in AFLOW-ICSD
\begin{table*}
  \begin{center}
    \caption{\textbf{Most frequent prototypes in the \AFLOW-\ICSD\ catalog.}
    \small
    The five most common prototypes are shown for unary, binary, ternary, and quaternary compounds as identified via \AFLOWXTALFINDERSHORT.
    The original and relaxed geometry sets are shown on the top and bottom portions of the table, respectively.
    Each prototype is listed with the following information: 
    \AFLOW\ label, number of unique atom decorations, representative compound with its \ICSD\ designation, 
    number of unique compounds exhibiting the structure (along with the count when including duplicate compounds), and matches to existing \AFLOW\ prototypes, if they exist.
    Empty rows in the \AFLOW\ prototype column reveal new prototypes, which will be included in Part 3 of the \AFLOW\ \PROTOTYPEENCYCLOPEDIA~\citeANRL.
    The complete list of prototypes is provided in the \SUPPLEMENTARYMATERIALS.
    }
    \resizebox{\linewidth}{!}{%
    \begin{tabular}{c l c l c l}
      & \multicolumn{1}{c}{\multirow{ 2}{*}{{\scriptsize{AFLOW}} label}} & \multicolumn{1}{c}{\multirow{ 2}{*}{\shortstack[c]{\# unique \\ decors.}}} & \multicolumn{1}{c}{\multirow{ 2}{*}{\shortstack[c]{representative \\ compd. (\scriptsize{ICSD} \#)}}} & \multicolumn{1}{c}{\multirow{2}{*}{\# compounds}} & \multicolumn{1}{c}{\multirow{2}{*}{\shortstack[c]{{\scriptsize{AFLOW}} prototype \\ (common name)}}} \\
      & & & & & \\
      \hline
      \hline
      \parbox[t]{2mm}{\multirow{20}{*}{\rotatebox[origin=c]{90}{original}}} & A\_cF4\_225\_a                      & 1   & Gd (20502)                                & 86 (379)       & 1, 2, \href{http://www.aflow.org/prototype-encyclopedia/A_cF4_225_a}{A\_cF4\_225\_a} (face-centered cubic, $A1$) \\
      & A\_cI2\_229\_a                      & 1   & H (28465)                                 & 58 (228)       & 58, 59, \href{http://www.aflow.org/prototype-encyclopedia/A_cI2_229_a}{A\_cI2\_229\_a} (body-centered cubic, $A2$) \\
      & A\_hP2\_194\_c                      & 1   & Be (1425)                                 & 54 (252)       & 115, 117, \href{http://www.aflow.org/prototype-encyclopedia/A_hP2_194_c}{A\_hP2\_194\_c-001} (hexagonal close packed, $A3$) \\
      & A\_cP1\_221\_a                      & 1   & Sb (52227)                                & 13 (16)        & \href{http://www.aflow.org/prototype-encyclopedia/A_cP1_221_a}{A\_cP1\_221\_a} ($\alpha$-Po, $A_{h}$) \\
      & A\_tI2\_139\_a                      & 1   & Ga (12174)                                & 11 (32)        & 303, 304, \href{http://www.aflow.org/prototype-encyclopedia/A_tI2_139_a.In}{A\_tI2\_139\_a-001} (In, $A6$) \\
      \cmidrule(l){2-6}
      & AB\_cP2\_221\_b\_a                  & 1   & ClCs (22173)                              & 429 (1026)     & 61, 1026, 1205, \href{http://www.aflow.org/prototype-encyclopedia/AB_cP2_221_b_a}{AB\_cP2\_221\_b\_a} (CsCl, $B2$) \\
      & AB\_cF8\_225\_b\_a                  & 1   & INa (44279)                               & 396 (2176)     & 201, 720, 1009, 1200, \href{http://www.aflow.org/prototype-encyclopedia/AB_cF8_225_a_b}{AB\_cF8\_225\_a\_b} (rocksalt, $B1$) \\
      & AB3\_cP4\_221\_a\_c                 & 2   & SiU$_{3}$ (1890)                          & 308 (905)      & 25, 26, \href{http://www.aflow.org/prototype-encyclopedia/AB3_cP4_221_a_c}{AB3\_cP4\_221\_a\_c} (Cu$_{3}$Au, $L1_{2}$) \\
      & A2B\_cF24\_227\_c\_b                & 2   & Al$_{2}$Ca (30213)                        & 251 (1258)     & 182, 183, 1042, \href{http://www.aflow.org/prototype-encyclopedia/A2B_cF24_227_d_a}{A2B\_cF24\_227\_d\_a} (cubic laves, $C15$) \\
      & AB\_cF8\_216\_a\_c                  & 1   & CuI (9098)                                & 124 (557)      & 218, 1007, 1201, \href{http://www.aflow.org/prototype-encyclopedia/AB_cF8_216_c_a}{AB\_cF8\_216\_c\_a} (zincblende, $B3$) \\
      \cmidrule(l){2-6}
      & A3BC\_cP5\_221\_c\_a\_b             & 6   & O$_{3}$PbTi (1613)                        & 414 (759)      & T0009, \href{http://www.aflow.org/prototype-encyclopedia/AB3C_cP5_221_a_c_b}{AB3C\_cP5\_221\_a\_c\_b} (cubic perovskite, $E2_{1}$) \\
      & A2BC\_cF16\_225\_c\_b\_a            & 3   & Cu$_{2}$LiSi (15128)                      & 293 (556)      & T0001, TBCC013,\href{http://www.aflow.org/prototype-encyclopedia/AB2C_cF16_225_a_c_b}{AB2C\_cF16\_225\_a\_c\_b} (Heusler, $L2_{1}$) \\
      & ABC\_hP9\_189\_f\_bc\_g             & 6   & RuSiZr (16306)                            & 227 (332)      & \\ % same family with ICSD 1133 (misfit=0.13734918807230)
      & ABC\_cF12\_216\_c\_a\_b             & 3   & AuMgSn (16475)                            & 188 (287)      & T0003, \href{http://www.aflow.org/prototype-encyclopedia/ABC_cF12_216_b_c_a}{ABC\_cF12\_216\_b\_c\_a} (half-Heusler, $C1_{b}$) \\
      & ABC\_oP12\_62\_c\_c\_c              & 6   & CoMoP (2421)                              & 185 (244)      & T0004 (CoGeMn ICSD:\#52968) \\
      \cmidrule(l){2-6}
      & A2BC6D\_cF40\_225\_c\_a\_e\_b       & 12  & Ba$_{2}$MnO$_{6}$W (189)                  & 207 (314)      & Q0001 (elpasolite) \\
      & ABCD\_tP8\_129\_b\_c\_a\_c          & 24  & AgLaOS (15530)                            & 54 (86)        & \\
      & AB3C7D\_hP24\_173\_a\_c\_b2c\_b     & 24  & CuLa$_{3}$S$_{7}$Si (23519)               & 50 (82)        & \\
      & A3BC6D\_hR22\_167\_e\_a\_f\_b       & 24  & Ca$_{3}$LiO$_{6}$Ru (50018)               & 49 (74)        & \\
      & A2B12C3D3\_cI160\_230\_a\_h\_d\_c   & 24  & Al$_{2}$F$_{12}$Li$_{3}$Na$_{3}$ (9923)   & 41 (219)       & \href{http://www.aflow.org/prototype-encyclopedia/A2B3C12D3_cI160_230_a_c_h_d}{A2B3C12D3\_cI160\_230\_a\_c\_h\_d-001} (garnet, $S1_{4}$) \\
      \hline
      \hline
      \parbox[t]{2mm}{\multirow{20}{*}{\rotatebox[origin=c]{90}{relaxed}}} & A\_cF4\_225\_a                      & 1   & Ce (2284)                                 & 68 (383)       & 1, 2, \href{http://www.aflow.org/prototype-encyclopedia/A_cF4_225_a}{A\_cF4\_225\_a} (face-centered cubic, $A1$) \\
      & A\_cI2\_229\_a                      & 1   & H (28465)                                 & 50 (231)       & 58, 59, \href{http://www.aflow.org/prototype-encyclopedia/A_cI2_229_a}{A\_cI2\_229\_a} (body-centered cubic, $A2$) \\
      & A\_hP2\_194\_c                      & 1   & Be (1425)                                 & 40 (244)       & 115, 117, \href{http://www.aflow.org/prototype-encyclopedia/A_hP2_194_c}{A\_hP2\_194\_c-001} (hexagonal close packed, $A3$) \\
      & A\_cP1\_221\_a                      & 1   & Sb (52227)                                & 12 (27)        & \href{http://www.aflow.org/prototype-encyclopedia/A_cP1_221_a}{A\_cP1\_221\_a} ($\alpha$-Po, $A_{h}$) \\
      & A\_tI2\_139\_a                      & 1   & Ga (12174)                                & 9 (31)         & 303, 304, \href{http://www.aflow.org/prototype-encyclopedia/A_tI2_139_a.In}{A\_tI2\_139\_a-001} (In, $A6$) \\
      \cmidrule(l){2-6}
      & AB\_cP2\_221\_a\_b                  & 1   & CsI (9204)                                & 399 (1041)     & 61, 1026, 1205, \href{http://www.aflow.org/prototype-encyclopedia/AB_cP2_221_b_a}{AB\_cP2\_221\_b\_a} (CsCl, $B2$) \\
      & AB\_cF8\_225\_b\_a                  & 1   & INa (44279)                               & 338 (2188)     & 201, 720, 1009, 1200, \href{http://www.aflow.org/prototype-encyclopedia/AB_cF8_225_a_b}{AB\_cF8\_225\_a\_b} (rocksalt, $B1$) \\
      & AB3\_cP4\_221\_a\_c                 & 2   & SiU$_{3}$ (1890)                          & 308 (915)      & 25, 26, \href{http://www.aflow.org/prototype-encyclopedia/AB3_cP4_221_a_c}{AB3\_cP4\_221\_a\_c} (Cu$_{3}$Au, $L1_{2}$) \\
      & A2B\_cF24\_227\_c\_b                & 2   & Fe$_{2}$Tb (2351)                         & 248 (1270)     & 182, 183, 1042, \href{http://www.aflow.org/prototype-encyclopedia/A2B_cF24_227_d_a}{A2B\_cF24\_227\_d\_a} (cubic laves, $C15$) \\
      & AB\_cF8\_216\_c\_a                  & 1   & CuI (9098)                                & 115 (557)      & 218, 1007, 1201, \href{http://www.aflow.org/prototype-encyclopedia/AB_cF8_216_c_a}{AB\_cF8\_216\_c\_a} (zincblende, $B3$) \\
      \cmidrule(l){2-6}
      & A3BC\_cP5\_221\_c\_a\_b             & 6   & O$_{3}$PbTi (1613)                        & 399 (841)      & T0009, \href{http://www.aflow.org/prototype-encyclopedia/AB3C_cP5_221_a_c_b}{AB3C\_cP5\_221\_a\_c\_b} (cubic perovskite, $E2_{1}$) \\
      & A2BC\_cF16\_225\_c\_b\_a            & 3   & Cu$_{2}$LiSi (15128)                      & 291 (556)      & T0001, TBCC013, \href{http://www.aflow.org/prototype-encyclopedia/AB2C_cF16_225_a_c_b}{AB2C\_cF16\_225\_a\_c\_b} (Heusler, $L2_{1}$) \\
      & AB2C2\_tI10\_139\_a\_e\_d           & 6   & CaGe$_{2}$Ni$_{2}$ (408)                  & 241 (572)      & T0011 (As$_{2}$CePd$_{2}$ ICSD:\#604354) \\
      & ABC\_hP9\_189\_f\_bc\_g             & 6   & RuSiZr (16306)                            & 230 (332)      & \\ 
      & ABC\_cF12\_216\_c\_b\_a             & 3   & AuMgSn (16475)                            & 188 (287)      & T0003, \href{http://www.aflow.org/prototype-encyclopedia/ABC_cF12_216_b_c_a}{ABC\_cF12\_216\_b\_c\_a} (half-Heusler, $C1_{b}$) \\
      \cmidrule(l){2-6}
      & A2BC6D\_cF40\_225\_c\_a\_e\_b       & 12  & Ba$_{2}$MnO$_{6}$W (189)                  & 209 (321)      & Q0001 (elpasolite) \\
      & ABCD\_tP8\_129\_b\_c\_a\_c          & 24  & AgLaOS (15530)                            & 55 (95)        & \\
      & A3BC6D\_hR22\_167\_e\_a\_f\_b       & 24  & Ca$_{3}$LiO$_{6}$Ru (50018)               & 45 (69)        & \\
      & AB3C7D\_hP24\_173\_a\_c\_b2c\_b     & 24  & CuLa$_{3}$S$_{7}$Si (23519)               & 40 (72)        & \\
      & A2B12C3D3\_cI160\_230\_a\_h\_d\_c   & 24  & Al$_{2}$F$_{12}$Li$_{3}$Na$_{3}$ (9923)   & 37 (218)       & \href{http://www.aflow.org/prototype-encyclopedia/A2B3C12D3_cI160_230_a_c_h_d}{A2B3C12D3\_cI160\_230\_a\_c\_h\_d-001} (garnet, $S1_{4}$) \\
      \hline
    \end{tabular}
    }
    \label{table:common_prototypes}
  \end{center}
\end{table*}
                                                      
While \AFLOW.org contains a subset of the \ICSD\ catalog, the highest frequency prototypes are consistent with  
those published for the \ICSD~\cite{Allmann_StructureTypesICSD_2007}.
In particular, \AFLOWXTALFINDERSHORT\ and the \ICSD\ both identify the following structures 
as some of the most common prototypes: 
Al$_{2}$MgO$_{4}$ (spinel, A2BC4\_cF56\_227\_d\_a\_e-001), 
CaTiO$_{3}$ (cubic perovskite, AB3C\_cP5\_221\_a\_c\_b), 
GdFeO$_{3}$ (AB3C\_oP20\_62\_a\_cd\_c), and 
NaCl (rocksalt, AB\_cF8\_225\_a\_b)  
(see Table~\ref{table:common_prototypes} and the \SUPPLEMENTARYMATERIALS).
The criteria for grouping compounds into structure-types described in Ref.~\cite{Allmann_StructureTypesICSD_2007} 
is more relaxed than \AFLOWXTALFINDERSHORT\ ({\it{e.g.}} larger tolerances for $c/a$ and $\beta$ ranges and user-defined ranges for fractional atomic coordinates). 
Consequently, \AFLOWXTALFINDERSHORT\ finds more distinct prototype structures than the 1,600 (as of January 2007) in Ref.~\cite{Allmann_StructureTypesICSD_2007}.

%– four structure types have more than 1000 representatives (Al2MgO4=spinel, CaTiO3=cubic perovskite, GdFeO3=AB3C\_oP20\_62\_c\_cd\_a, and NaCl=rocksalt),
%– 13 structure types have 500–999 representatives: (AuCu3=L1_2, CeAl3Ga2, CsCl, Cu=fcc, Cu2Mg=cubic_laves, K2MgF4, Mg2SiO4, MgSrSi, NaCrS2, NdAlO3, PbCl2, PbClF and YbBa2Cu3O6+x(orh),

From this analysis, new candidate prototypes have been identified that are missing from the \PROTOTYPEENCYCLOPEDIA\ 
(signified by empty rows in the last columns of Table~\ref{table:common_prototypes} and the \SUPPLEMENTARYMATERIALS).
The number of new prototypes in the original (relaxed) sets with more than 
10 unique compounds exhibiting the structure are: 
binaries 31 (33),
ternaries 168 (177),
quaternaries 40 (42), and
quinaries 4 (3);
while the unaries, senaries, and septenaries have 0 (0).
This amounts to 243 distinct crystalline structures that will be incorporated 
into future installments of the \PROTOTYPEENCYCLOPEDIA{} ~\footnote{Most entries in the Prototype Encyclopedia stem from 
experimentally observed structures; therefore, we plan to use the original geometries for prototyping.}. 

Some structures in Table~\ref{table:common_prototypes} and the \SUPPLEMENTARYMATERIALS\ are equivalent to 
the \PROTOTYPEENCYCLOPEDIA\ prototypes with a different number of atom types. 
For example, the third most common ternary ABC\_hP9\_189\_f\_bc\_g (RuSiZr \ICSD\ \#16306, original geometry) 
matches to the binary analog A2B\_hP9\_189\_fg\_bc (Fe$_{2}$P, {\it{Strukturbericht}}: $C22$)~\cite{A2B_hP9_189_fg_bc}
when the $f$ and $g$ Wyckoff positions are of the same atom type. %, consistent with reports from Villars~\cite{}.
We classify the prototypes as distinct; similar to distinguishing between the diamond ($n=1$) and zincblende ($n=2$) structures.

%\section*{Discussion} \label{sec:conclusion}
Herein, we present \AFLOWXTALFINDERSHORT: a software for automatically
identifying unique prototypes and calculating structural similarity of crystals.
The framework performs robust symmetry, local atomic geometry, and geometric structure 
comparisons.
Routines are available to quantify structural similarity for 
\textbf{i.} compounds (material-type comparisons),
\textbf{ii.} prototypes (structure-type),
\textbf{iii.} atom decorations (decoration-type), and
\textbf{iv.} spin configurations (magnetic-type).
The program can analyze multiple structures 
simultaneously and aggregate them into equivalent groups, with multithreading capabilities available for improving performance.
Built-in methods compare input structures to the \AFLOW.org repository and the 
\AFLOW\ prototype libraries for detecting new compounds and structure-types.
Crystal prototyping techniques are also introduced to cast structures into a standard designation, facilitating extensions of the \PROTOTYPEENCYCLOPEDIA.
A command line and Python interface are provided for easing incorporation into user-workflows.
Applying the procedures to the \AFLOW-\ICSD\ repository revealed approximately 
15,000 prototypes out of over 60,000 \ICSD\ entries, representing over 34,000 unique compounds.
Subsequent comparisons with the \AFLOW\ prototype libraries exposed new candidate entries  
for future iterations of the encyclopedia.
%Overall, \AFLOWXTALFINDERSHORT\ serves as a versatile tool for comparing crystal structures and 
%generating prototypes.
Overall, \AFLOWXTALFINDERSHORT\ serves as a versatile tool for finding prototypes and 
comparing crystalline geometries.

%%%%%%%%%%%%%%%%%%%%%%%%%%%%%%%%%%%%%%%%%%%%%%%%%%%%%%%%%%%%%%%%%%%%%%%%%%%%%%%%%%%%%%%%%%%%%%%%%%%%%%%%%%%%%%%%%%%%%%
%
% START: Methods
%
%%%%%%%%%%%%%%%%%%%%%%%%%%%%%%%%%%%%%%%%%%%%%%%%%%%%%%%%%%%%%%%%%%%%%%%%%%%%%%%%%%%%%%%%%%%%%%%%%%%%%%%%%%%%%%%%%%%%%%

%\subsection*{Command-line interface}                     
\label{subsec:command_line}                           
\noindent\textbf{Command-line interface.} 
The \AFLOWXTALFINDERSHORT\ command-line calls are detailed below.
Function descriptions and options are provided following each command.

\noindent\textbf{Prototype commands.}
\begin{myitemize}                       
  \item \verbWrap!aflow --prototype < file!             
    \begin{myitemize}                                   
      \item{Converts a structure (\verbWrap!file!) into its standard \AFLOW\ prototype label. 
        The parameter variables (degrees of freedom) and corresponding values are also listed.
        Information about the label and parameters are described in the Refs.\citeANRL. \\
        Options specific to this command:
        \begin{myitemize_nobullet}
            \item{\verbWrap!--setting=1|2|aflow! 
                \begin{myitemize}
                \item{
                  Specify the space group setting for the conventional cell/Wyckoff positions.  
                  The \verbWrap!aflow! setting follows the choices of the \PROTOTYPEENCYCLOPEDIA: 
                  axis-$b$ for monoclinic space groups, 
                  rhombohedral setting for rhombohedral space groups, 
                  and origin centered on the inversion site for centrosymmetric space groups (default: \verbWrap!aflow!).
              }
                \end{myitemize}
        }
        \end{myitemize_nobullet}
    }
  \end{myitemize}
  \item \verbWrapBlue!aflow --proto=<label>.<ABC..>:Ag:C:Cu:... --params=parameter_1,parameter_2,...!             
    \begin{myitemize}                                   
      \item{
        {{
        Generates a geometry file based on the ideal prototype designation (\verbWrapBlue!label!) 
        and parameter values (\verbWrapBlue!parameter_1,parameter_2,...!).
        A particular atom decoration can be specified after the label (\verbWrapBlue!<ABC...>!). 
        By default, the structure is created with fictitious atoms ({\it{i.e.}} A, B, C, D, ...); however,
        this can be overwritten by appending real elements to the label separated by colons 
        ({\it{e.g.}} \verbWrapBlue!<label>.<ABC...>:Ag:C:Cu:...!).
        Options specific to this command:
        }}
        \begin{myitemize_nobullet}
            \item{\verbWrapBlue!--add_equations! 
                \begin{myitemize}
                \item{
                  {{
                  The symbolic version of the geometry file (in terms of the variable degrees of freedom) 
                  is printed after the numeric geometry file.
                  }}
                }
                \end{myitemize}
            }
            \item{\verbWrapBlue!--equations_only! 
                \begin{myitemize}
                \item{
                  {{
                  Only print the symbolic version of the geometry file (in terms of the variable degrees of freedom).
                  }} 
                }
                \end{myitemize}
            }
        \end{myitemize_nobullet}
    }
  \end{myitemize}
\end{myitemize}
\noindent\textbf{Comparison commands.}
\begin{myitemize}                       
  \item \verbWrap!aflow --compare_materials!
  \begin{myitemize}
    \item{Compares compounds comprised of the same atomic species and with 
        commensurate stoichiometric ratios, {\it{i.e.}} material-type comparison, and returns their level of similarity (misfit value).
        This method identifies unique and duplicate materials.
        There are three input types:
        \begin{myitemize}
            \item{\verb!aflow --compare_materials=<f1>,<f2>,...!: append geometry files (\verbWrap!<f1>,<f2>,...!) to compare,}
            \item{\verb!aflow --compare_materials -D <path>!: specify path to directory (\verbWrap!<path>!) containing geometry files to compare, and}
            \item{\verb!aflow --compare_materials -F=<filename>!: specify file (\verbWrap!<filename>!) containing compounds between delimiters \verb![VASP_POSCAR_MODE_EXPLICIT]START! and \verb![VASP_POSCAR_MODE_EXPLICIT]STOP!.
                Additional delimiters will be included in later versions.}
        \end{myitemize}
  }
  \end{myitemize}
  \item \verbWrap!aflow --compare_structures!
  \begin{myitemize}
    \item{Compares compounds with commensurate stoichiometric ratios with no 
        requirement of the atomic species, {\it{i.e.}} structure-type comparison, and returns their level of similarity (misfit value).
        This method identifies unique and duplicate prototypes.
        There are three input types:
        \begin{myitemize}
            \item{\verb!aflow --compare_structures=<f1>,<f2>,...!: append geometry files (\verbWrap!<f1>,<f2>,...!) to compare,}
            \item{\verb!aflow --compare_structures -D <path>!: specify path to directory (\verbWrap!<path>!) containing geometry files to compare, and}
            \item{\verb!aflow --compare_structures -F=<filename>!: specify file (\verbWrap!<filename>!) containing compounds between delimiters \verb![VASP_POSCAR_MODE_EXPLICIT]START! and \verb![VASP_POSCAR_MODE_EXPLICIT]STOP!.
                Additional delimiters will be included in later versions.}
        \end{myitemize}
  }
  \end{myitemize}
  \item \verbWrap!aflow --compare2database < file!
    \begin{myitemize}
      \item{
           Compares a structure (\verbWrap!file!) to \AFLOW\ database entries, returning similar 
           compounds and quantifying their levels of similarity (misfit values).
           Material properties can be extracted from the database (via \AFLUX) and printed, highlighting structure-property relationships.
           Performs material-type comparisons or structure-type comparisons (by adding the \verbWrap!--structure_comparison! option).
           Options specific to this command:
           \begin{myitemize_nobullet}
            \item{\verbWrap!--properties=<keyword,keyword,...>!
                \begin{myitemize}
                    \item{Specify the properties via their \API\ keyword to print the corresponding values with the comparison results.}
                \end{myitemize}
            }
            \item{\verbWrap!--catalog=<string>! 
                \begin{myitemize}
                    \item{Restrict the database entries to a specific catalog/library ({\it{e.g.}} `lib1', `lib2', `lib3', `icsd', etc.).}
                \end{myitemize}
            }
            \item{\verbWrap!--geometry_file=<string>!  
                \begin{myitemize}
                    \item{Compare geometries from a particular DFT relaxation step ({\it{e.g.}} `POSCAR.relax1', `POSCAR.relax2', `POSCAR.static', etc.).}
                \end{myitemize}
            }
           \end{myitemize_nobullet}
      }
  \end{myitemize}
  \item \verbWrap!aflow --compare2prototypes < file!
    \begin{myitemize}
      \item{
           Compares a structure (\verbWrap!file!) against the \AFLOW\ prototype libraries, returning similar 
           structures and quantifying their levels of similarity (misfit values).
           \begin{myitemize_nobullet}
            \item{\verbWrap!--catalog=<string>! 
                \begin{myitemize}
                    \item{Restrict the prototypes to a specific catalog/library ({\it{e.g.}} `aflow' or `htqc').}
                \end{myitemize}
            }
           \end{myitemize_nobullet}
      }
  \end{myitemize}
  \item \verbWrap!aflow --isopointal_prototypes < file!
    \begin{myitemize}
      \item{
           Returns prototype labels that are isopointal ({\it{i.e.}} similar space group and Wyckoff positions) 
           to the input structure (\verbWrap!file!).
           \begin{myitemize_nobullet}
            \item{\verbWrap!--catalog=<string>! 
                \begin{myitemize}
                    \item{Restrict the prototypes to a specific catalog/library ({\it{e.g.}} `aflow' or `htqc').}
                \end{myitemize}
            }
           \end{myitemize_nobullet}
      }
  \end{myitemize}
  \item \verbWrap!aflow --unique_atom_decorations < file!
    \begin{myitemize}
      \item{
         Determines the unique and duplicate atom decorations for a given structure.
      }
    \end{myitemize}
\end{myitemize}

% options
Generic options for all comparison commands (unless indicated otherwise):
  \begin{myitemize}
  \item \verbWrapBlue!--misfit_match=<number>|! \\
    \verbWrapBlue!--misfit_match_threshold=<number>!
    \begin{myitemize}
      \item{{{Specifies the misfit threshold for matched structures (default: $\epsilon_{\mathrm{match}}=0.1$).}}}
    \end{myitemize}
  \item \verbWrapBlue!--misfit_family=<number>|! \\
    \verbWrapBlue!--misfit_family_threshold=<number>!
    \begin{myitemize}
      \item{{{Specifies the misfit threshold for structures in the ``same family'' (default: $\epsilon_{\mathrm{family}}=0.2$).}}}
    \end{myitemize}
  \item \verbWrap!--np=<number>|--num_proc=<number>!
    \begin{myitemize}
      \item{Allocate the number of processors/threads for the task.}
    \end{myitemize}
  \item \verbWrap!--optimize_match!
    \begin{myitemize}
      \item{Explore all lattice and origin choices to find the best matching representation, {\it{i.e.}} minimizes misfit value.}
    \end{myitemize}
  \item \verbWrap!--no_scale_volume!
    \begin{myitemize}
      \item{Suppresses volume rescaling during structure matching; identifies differences due to volume expansion or compression of a structure.}
    \end{myitemize}
  \item \verbWrap!--ignore_symmetry!
    \begin{myitemize}
      \item{Neglects symmetry (both space group and Wyckoff positions) for grouping comparisons.}
    \end{myitemize}
  \item \verbWrap!--ignore_Wyckoff!
    \begin{myitemize}
      \item{Neglects Wyckoff symmetry (site symmetry) for filtering comparisons, but considers the space group number.}
    \end{myitemize}
  \item \verbWrap!--ignore_local_geometry!
    \begin{myitemize}
      \item{Neglects local LFA geometries for filtering comparisons.}
    \end{myitemize}
  \item \verbWrap!--minkowski!
    \begin{myitemize}
      \item{Performs a Minkowski lattice transformation~\cite{MinkowskiReduction} on all structures prior to comparison; offering a speed increase.}
    \end{myitemize}
  \item \verbWrap!--niggli!
    \begin{myitemize}
      \item{Performs a Niggli lattice transformation~\cite{Niggli1928} on all structures prior to comparison; offering a speed increase.}
    \end{myitemize}
  \item \verbWrap!--primitive|--primitivize!
    \begin{myitemize}
      \item{Converts all structures to a primitive form prior to comparison; offering a speed increase.}
    \end{myitemize}
  \item \verbWrap!--keep_unmatched!
    \begin{myitemize}
      \item{Retains misfit information of unmatched structures ({\it{i.e.}} {
      {$\epsilon>\epsilon_{\mathrm{match}}$}}).}
    \end{myitemize}
  \item \verbWrap!--match_to_aflow_prototypes!
    \begin{myitemize}
        \item{Identifies matching \AFLOW\ prototypes to the representative structure.  The option does not apply to \verbWrap!--unique_atom_decorations! or \verbWrap!--compare2prototypes! (redundant).}
    \end{myitemize}
  \item \verbWrap!--magmom=<m1,m2,...|INCAR|OUTCAR>:...!
    \begin{myitemize}
        \item{Specifies the magnetic moment for each structure (collinear or non-collinear) delimited by colons, signaling a magnetic-type comparison. 
            The option does not apply to \verbWrap!--compare_structures! since the atom type is neglected.
            \AFLOWXTALFINDERSHORT\ supports three input formats for the magnetic moment: 
            \textbf{i.} explicit declaration via comma-separated string $m_{1},m_{2},...m_{n}$ ($m_{1,x},m_{1,y},m_{1,z},m_{2,x},...m_{n,z}$ for non-collinear) 
            \textbf{ii.} read from a \VASP\ \INCAR, or
            \textbf{iii.} read from a \VASP\ \OUTCAR.
            Additional magnetic moment readers for other {\it{ab initio}} codes will be available in future versions.}
    \end{myitemize}
  \item \verbWrap!--add_aflow_prototype_designation!
    \begin{myitemize}
        \item{Casts representative structure into the \AFLOW\ standard designation.  The option does not apply to commands \verbWrap!--unique_atom_decorations! or \verbWrap!--prototype! (redundant).}
    \end{myitemize}
  \item \verbWrap!--remove_duplicate_compounds!
    \begin{myitemize}
        \item{For structure-type comparisons, duplicate compounds are identified first (via a material-type comparison without volume scaling), then remaining unique compounds are compared, removing duplicate bias.}
    \end{myitemize}
  \item \verbWrap!--print!
    \begin{myitemize}
        \item{For comparing two structures, additional comparison information is printed, including atom mappings, distances between matched atoms, and the transformed structures in the closest matching representation.}
    \end{myitemize}
  \item \verbWrap!--print=text|json!
    \begin{myitemize}
        \item{For comparing multiple structures, the results are printed to into human-readable text or \JSON\ files, respectively.  By default, \AFLOWXTALFINDERSHORT\ writes the output to both files.}
    \end{myitemize}
  \item \verbWrap!--quiet!
    \begin{myitemize}
        \item{Suppresses the log information for the comparisons.}
    \end{myitemize}
  \item \verbWrap!--screen_only!
    \begin{myitemize}
        \item{Prints the comparison results to the screen and does not write to any files.}
    \end{myitemize}
  \end{myitemize}

\noindent\textbf{Python environment.} In addition to the command-line interface, a Python module is available 
for inclusion into a variety of workflows.  
The module mirrors the format used for \AFLOWSYM~\cite{curtarolo:art135} and \AFLOWCHULL~\cite{curtarolo:art144}.
An \AFLOWXTALFINDERSHORT\ function is performed on the input(s) and the results are returned to an \verbWrap!XtalFinder!~class. 
The module wraps around a local instance of \AFLOW, and the path to the \AFLOW\ executable can be specified by: \\
\verbWrap!XtalFinder(aflow_executable=`your_executable')!. \\
By default, the \verbWrap!XtalFinder! object searches for an \AFLOW\ executable in the \verbWrap!PATH!.
An example Python script is shown below, where 
\verbWrap!XtalFinder! object is initialized and a material-type comparison 
between two structure files (\POSCAR{}s) is performed.
%DO NOT CHANGE INDENT AS IT IS VERBATIM 
\begin{python}
from aflow_xtal_match import XtalFinder
from pprint import pprint 

xtal_match = XtalFinder(aflow_executable='./aflow')
input_files = [`test1.poscar',`test2.poscar']
output = xtal_match.compare_materials(input_files)
pprint(output)
\end{python}
%DO NOT CHANGE INDENT AS IT IS VERBATIM ARGHHHH

The following Python functions are accessible, corresponding to the commands described in the previous section:
\begin{myitemize}
  \item \verbWrap|get_prototype_label(input_file, options)|
  \item \verbWrap|compare_materials(input_files, options)|
  \item \verbWrap|compare_materials_directory(directory, options)|
  \item \verbWrap|compare_materials_file(filename, options)|
  \item \verbWrap|compare_structures(input_files, options)|
  \item \verbWrap|compare_structures_directory(directory, options)|
  \item \verbWrap|compare_structures_file(filename, options)|
  \item \verbWrap|compare2database(input_file, options)|
  \item \verbWrap|compare2prototypes(input_file, options)|
  \item \verbWrap|get_isopointal_prototypes(input_file, options)|
  \item \verbWrap|get_unique_atom_decorations(input_file, options)|
\end{myitemize}
The input fields for the Python functions are as follows:
\begin{myitemize}
  \item{\verbWrap!input_file!}
  \begin{myitemize}
    \item{A string specifying the path to a structure file,
        {\it{e.g.}} \verbWrap!input_file=`/home/user/test.poscar'!.}
  \end{myitemize}
  \item{\verbWrap!input_files!}
  \begin{myitemize}
    \item{A list of paths (of any size $\ge 2$) to structure files,
        {\it{e.g.}} \verbWrap!input_files=[`test1.poscar', ...]!.}
  \end{myitemize}
  \item{\verbWrap!directory!}
  \begin{myitemize}
    \item{A string specifying the path to directory containing structure files,
        {\it{e.g.}} \verbWrap!directory=`/home/user/directory'!.}
  \end{myitemize}
  \item{\verbWrap!filename!}
  \begin{myitemize}
    \item{A string specifying the path to a file containing structure files separated by a delimiter,
        {\it{e.g.}} \verbWrap!filename=`/home/user/list_of_structures.txt'!.}
  \end{myitemize}
  \item{\verbWrap!options!}
  \begin{myitemize}
      \item{A string specifying non-default functionality (optional), which 
          has the form \verbWrap!--<flag>! or \verbWrap!--<keyword>=<value>!,
        {\it{e.g.}} ``\verbWrap!--ignore_symmetry --np=8!''.}
  \end{myitemize}
\end{myitemize}

%\section*{Python module}
\label{sec:xtal_match_python_module}
\textbf{Python module.} Below is a Python module for the \AFLOWXTALFINDERSHORT\ functionality.
All output is converted into \underline{J}ava\underline{S}cript \underline{O}bject \underline{N}otation (\JSON) to ease 
integration into user workflows.
%DO NOT CHANGE INDENT AS IT IS VERBATIM ARGHHHH
\begin{python}
import json
import subprocess
import os

class XtalFinder:

    def __init__(self, aflow_executable='aflow'):
        self.aflow_executable = aflow_executable

    def aflow_command(self, cmd):
        try:
            return subprocess.check_output(
                self.aflow_executable + cmd,
                shell=True
            )
        except subprocess.CalledProcessError:
            print "Error aflow executable not found at: " + self.aflow_executable
    
    def get_prototype_label(self, input_file, options=None):
        fpath = os.path.realpath(input_file)
        command = ' --prototype'
        output = ''
        
        if options:
            command += ' ' + options
        
        output = self.aflow_command(
            command + ' --print=json < ' + fpath 
        )
        
        res_json = json.loads(output)
        return res_json

    def compare_materials(self, input_files, options=None):
        command = ' --compare_materials=' + ','.join(input_files)
        output = ''
        
        if options:
            command += ' ' + options
        
        output = self.aflow_command(
            command + ' --print=json --screen_only --quiet'
        )
        
        res_json = json.loads(output)
        return res_json

    def compare_materials_directory(self, directory, options=None):
        command = ' --compare_materials -D ' + directory
        output = ''
        
        if options:
            command += ' ' + options
        
        output = self.aflow_command(
            command + ' --print=json --screen_only --quiet'
        )
        
        res_json = json.loads(output)
        return res_json
    
    def compare_materials_file(self, filename, options=None):
        command = ' --compare_materials -F=' + filename
        output = ''
        
        if options:
            command += ' ' + options
        
        output = self.aflow_command(
            command + ' --print=json --screen_only --quiet'
        )
        
        res_json = json.loads(output)
        return res_json
    
    def compare_structures(self, input_files, options=None):
        command = ' --compare_structures=' + ','.join(input_files)
        output = ''
        
        if options:
            command += ' ' + options
        
        output = self.aflow_command(
            command + ' --print=json --screen_only --quiet'
        )
        
        res_json = json.loads(output)
        return res_json

    def compare_structures_directory(self, directory, options=None):
        command = ' --compare_structures -D ' + directory
        output = ''
        
        if options:
            command += ' ' + options
        
        output = self.aflow_command(
            command + ' --print=json --screen_only --quiet'
        )
        
        res_json = json.loads(output)
        return res_json
    
    def compare_structures_file(self, filename, options=None):
        command = ' --compare_structures -F=' + filename
        output = ''
        
        if options:
            command += ' ' + options
        
        output = self.aflow_command(
            command + ' --print=json --screen_only --quiet'
        )
        
        res_json = json.loads(output)
        return res_json
    
    def compare2database(self, input_file, options=None):
        fpath = os.path.realpath(input_file)
        command = ' --compare2database'
        output = ''
        
        if options:
            command += ' ' + options
        
        output = self.aflow_command(
            command + ' --print=json --screen_only --quiet < ' + fpath 
        )
        
        res_json = json.loads(output)
        return res_json

    def compare2prototypes(self, input_file, options=None):
        fpath = os.path.realpath(input_file)
        command = ' --compare2prototypes'
        output = ''
        
        if options:
            command += ' ' + options
        
        output = self.aflow_command(
            command + ' --print=json --screen_only --quiet < ' + fpath 
        )
        
        res_json = json.loads(output)
        return res_json
    
    def get_isopointal_prototypes(self, input_file, options=None):
        fpath = os.path.realpath(input_file)
        command = ' --isopointal_prototype'
        output = ''
        
        if options:
            command += ' ' + options
        
        output = self.aflow_command(
            command + ' --print=json < ' + fpath 
        )
        
        res_json = json.loads(output)
        return res_json
    
    def get_unique_atom_decorations(self, input_file, options=None):
        fpath = os.path.realpath(input_file)
        command = ' --unique_atom_decorations'
        output = ''
        
        if options:
            command += ' ' + options
        
        output = self.aflow_command(
            command + ' --print=json < ' + fpath 
        )
        
        res_json = json.loads(output)
        return res_json

\end{python}
%DO NOT CHANGE INDENT AS IT IS VERBATIM ARGHHHH

\label{subsec:output_details}
\noindent\textbf{\AFLOWXTALFINDER\ \JSON\ output details.} The output keywords for the \AFLOWXTALFINDERSHORT\ functions are listed below as they 
appear in the \JSON\ format.
The output for multiple comparisons (user defined sets, comparison to \AFLOW\ prototypes, and 
comparison to \AFLOW.org entries), unique atom decorations, and casting into the 
\AFLOW\ prototype representation are described.

\def\description{\item {{\it Description:}\ }}
\def\type{\item {{\it Type:}\ }}
\def\similarto{\item {{\it Similar to:}\ }}
\def\bluedescription{{{\item {{\it Description:}\ }}}} %
\def\bluetype{{{\item {{\it Type:}\ }}}}               %
\def\bluesimilarto{{{\item {{\it Similar to:}\ }}}}    %
\noindent\textbf{AFLOW prototype designation.}
\begin{myitemize}
\item \verbWrap|aflow_prototype_label|
  \begin{myitemize}
    \description \AFLOW\ label for the structure.
    \type \verbWrap|string|
  \end{myitemize}
\item \verbWrap|aflow_prototype_params_list|
  \begin{myitemize}
    \description degrees of freedom (variables) in the lattice and/or Wyckoff positions for the structure.  
    \type \verbWrap|array of strings|
  \end{myitemize}
\item \verbWrap|aflow_prototype_params_values|
  \begin{myitemize}
    \description values specifying the degrees of freedom for the structure.  
    \type \verbWrap|array of floats|
  \end{myitemize}
\end{myitemize}

\noindent\textbf{Comparison results.}
\begin{myitemize}
\item \verbWrap|structure_representative|
  \begin{myitemize}
    \description name of the representative structure for the prototype structure.
    \type \verbWrap|string|
  \end{myitemize}
\item \verbWrap|stoichiometry|
  \begin{myitemize}
    \description stoichiometry of the prototype structure.
    \type \verbWrap|array of integers|
  \end{myitemize}
\item \verbWrap|number_of_types|
  \begin{myitemize}
    \description number of atom types (species) in the prototype structure.
    \type \verbWrap|integer|
  \end{myitemize}
\item \verbWrap|number_of_atoms|
  \begin{myitemize}
    \description number of atoms in the unit cell (from the representative structure).
    \type \verbWrap|integer|
  \end{myitemize}
\item \verbWrap|elements|
  \begin{myitemize}
    \description atomic elements found in this structure from both the representative and duplicate compounds/structures.
    \type \verbWrap|array of strings|
  \end{myitemize}
\item \verbWrap|space_group|
  \begin{myitemize}
    \description space group number for the prototype structure.
    \type \verbWrap|integer|
  \end{myitemize}
\item \verbWrap|grouped_Wyckoff_positions|
  \begin{myitemize}
    \description Wyckoff positions grouped by atomic species (corresponding to the representative structure).
    \type \verbWrap|array of Wyckoff objects|
  \end{myitemize}
\item \verbWrap|geomeries_LFA|
  \begin{myitemize}
    \description local atomic geometries comprised of LFA types only (corresponding to the representative structure). 
    \type \verbWrap|array of local_geometry objects|
  \end{myitemize}
\item \verbWrap|property_names|
  \begin{myitemize}
    \description \API\ keywords corresponding to material properties (available for comparisons to the \AFLOW.org repository only).
    \type \verbWrap|array of strings|
  \end{myitemize}
\item \verbWrap|property_units|
  \begin{myitemize}
    \description units, if applicable, for material properties (available for comparisons to the \AFLOW.org repository only).
    \type \verbWrap|array of strings|
  \end{myitemize}
\item \verbWrap|structures_duplicate|
  \begin{myitemize}
    \description names of duplicate structures that match with the representative structure, {\it{i.e.}} misfit is less than 
    {{$\epsilon_{\mathrm{match}}$}}.
    \type \verbWrap|array of strings|
  \end{myitemize}
\item \verbWrap|misfits_duplicate|
  \begin{myitemize}
    \description values of the misfit between the representative structure and the duplicate structures.
    \type \verbWrap|array of floats|
  \end{myitemize}
\item \verbWrap|lattice_deviations_duplicate|
  \begin{myitemize}
    \description values of the lattice deviation between the representative structure and the duplicate structures.
    \type \verbWrap|array of floats|
  \end{myitemize}
\item \verbWrap|coordinate_displacements_duplicate|
  \begin{myitemize}
    \description values of the coordinate displacement between the representative structure and the duplicate structures.
    \type \verbWrap|array of floats|
  \end{myitemize}
\item \verbWrap|failures_duplicate|
  \begin{myitemize}
    \description values of the figure of failure between the representative structure and the duplicate structures.
    \type \verbWrap|array of floats|
  \end{myitemize}
\item \verbWrap|structures_family|
  \begin{myitemize}
    \description names of structures that are within the same family as the representative structure, {\it{i.e.}} misfit is between {{$\epsilon_{\mathrm{match}}$ and $\epsilon_{\mathrm{family}}$}}.
    \type \verbWrap|array of strings|
  \end{myitemize}
\item \verbWrap|misfits_family|
  \begin{myitemize}
    \description values of the misfit between the representative structure and the same family structures.
    \type \verbWrap|array of floats|
  \end{myitemize}
\item \verbWrap|lattice_deviations_family|
  \begin{myitemize}
    \description values of the lattice deviation between the representative structure and the same family structures.
    \type \verbWrap|array of floats|
  \end{myitemize}
\item \verbWrap|coordinate_displacements_family|
  \begin{myitemize}
    \description values of the coordinate displacement between the representative structure and the same family structures.
    \type \verbWrap|array of floats|
  \end{myitemize}
\item \verbWrap|failures_family|
  \begin{myitemize}
    \description values of the figure of failure between the representative structure and the same family structures.
    \type \verbWrap|array of floats|
  \end{myitemize}
\item \verbWrap|properties_structure_representative|
  \begin{myitemize}
    \description values of the material properties requested for the representative structure (available for comparisons to the \AFLOW.org repository only).
    \type \verbWrap|array of strings|
  \end{myitemize}
\item \verbWrap|properties_structures_duplicate|
  \begin{myitemize}
    \description values of the material properties requested for the duplicate structures (available for comparisons to the \AFLOW.org repository only).
    \type \verbWrap|2D array of strings|
  \end{myitemize}
\item \verbWrap|properties_structures_family|
  \begin{myitemize}
    \description values of the material properties requested for the same family structures (available for comparisons to the \AFLOW.org repository only).
    \type \verbWrap|2D array of strings|
  \end{myitemize}
\item \verbWrap|number_compounds_matching_representative|
  \begin{myitemize}
    \description number of compounds that match with the representative structure via a material-type comparison (only for structure-type comparisons that remove duplicate compounds beforehand).
    \type \verbWrap|integer|
  \end{myitemize}
\item \verbWrap|number_compounds_matching_duplicate|
  \begin{myitemize}
    \description number of compounds that match with the duplicates structures via a material-type comparison (only for structure-type comparisons that remove duplicate compounds beforehand).
    \type \verbWrap|array of integers|
  \end{myitemize}
\item \verbWrap|number_compounds_matching_family|
  \begin{myitemize}
    \description number of compounds that match with the same family structures via a material-type comparison (only for structure-type comparisons that remove duplicate compounds beforehand).
    \type \verbWrap|array of integers|
  \end{myitemize}
\item \verbWrap|matching_aflow_prototypes|
  \begin{myitemize}
    \description labels of \AFLOW\ crystal prototypes~\citeANRL\ that match with this structure (included when using option \\ ``\verbWrap|--add_matching_aflow_prototypes|''). 
    \type \verbWrap|array of strings|
  \end{myitemize}
\end{myitemize}
A \verbWrap!Wyckoff! object contains the following:
\begin{myitemize}
  \item \verbWrap|element|
  \begin{myitemize}
    \description atomic species on Wyckoff site. 
    \type \verbWrap|string|
  \end{myitemize}
  \item \verbWrap|type|
  \begin{myitemize}
    \description an index corresponding to atomic species, based on alphabetic ordering of element name. 
    \type \verbWrap|integer|
  \end{myitemize}
  \item \verbWrap|letters|
  \begin{myitemize}
    \description Wyckoff letters for the atomic species. 
    \type \verbWrap|array of strings|
  \end{myitemize}
  \item \verbWrap|multiplicities|
  \begin{myitemize}
    \description Wyckoff multiplicities for the atomic species. 
    \type \verbWrap|array of integers|
  \end{myitemize}
  \item \verbWrap|site_symmetries|
  \begin{myitemize}
    \description Wyckoff site symmetries for the atomic species. 
    \type \verbWrap|array of strings|
  \end{myitemize}
\end{myitemize}
A \verbWrap!local_geometry! object contains the following:
\begin{myitemize}
  \item \verbWrap|center_element|
  \begin{myitemize}
    \description atomic species at the center of the geometry cluster.
    \type \verbWrap|string|
  \end{myitemize}
  \item \verbWrap|center_type|
  \begin{myitemize}
    \description index corresponding to atomic species at the center of the geometry cluster; enumeration is based on alphabetic ordering of element name. 
    \type \verbWrap|integer|
  \end{myitemize}
  \item \verbWrap|neighbor_elements|
  \begin{myitemize}
    \description atomic elements of neighbors. 
    \type \verbWrap|array of strings|
  \end{myitemize}
  \item \verbWrap|neighbor_distances|
  \begin{myitemize}
    \description distances of the neighbors from the center atom
    \type \verbWrap|array of floats|
  \end{myitemize}
  \item \verbWrap|neighbor_frequencies|
  \begin{myitemize}
    \description coordination of the neighbors at the corresponding neighbor distance (within 10\%).
    \type \verbWrap|array of integers|
  \end{myitemize}
  \item \verbWrap|neighbor_coordinates|
  \begin{myitemize}
    \description coordinates of the neighbors that comprise the local atomic geometry; the origin of the system resides on the center atom.
    \type \verbWrap|2D array of floats|
  \end{myitemize}
\end{myitemize}

\noindent\textbf{Permutation results.}
\begin{myitemize}
\item \verbWrap|atom_decorations_equivalent|
  \begin{myitemize}
    \description groupings of equivalent atom decorations for the structure.  
    \type \verbWrap|2D array of strings|
  \end{myitemize}
\end{myitemize}

\noindent\textbf{{{Ideal prototype \API\ keywords.}}}
\begin{myitemize}
\item \verbWrapBlue|aflow_prototype_label_orig|
  \begin{myitemize}
    \bluedescription {{the standard prototype label of the structure (original geometry).}}
    \bluetype \verbWrapBlue|string|
  \end{myitemize}
\item \verbWrapBlue|aflow_prototype_params_list_orig|
  \begin{myitemize}
    \bluedescription {{ degrees of freedom (variables) in the lattice and/or Wyckoff positions of the structure (original geometry).}}
    \bluetype \verbWrapBlue|array of strings|
  \end{myitemize}
\item \verbWrapBlue|aflow_prototype_params_values_orig|
  \begin{myitemize}
    \bluedescription {{ values specifying the degrees of freedom of the structure (original geometry).}}
    \bluetype \verbWrapBlue|array of floats|
  \end{myitemize}
\item \verbWrapBlue|aflow_prototype_label_relax|
  \begin{myitemize}
    \bluedescription {{ the standard prototype label of the structure (DFT-relaxed geometry).}}
    \bluetype \verbWrapBlue|string|
  \end{myitemize}
\item \verbWrapBlue|aflow_prototype_params_list_relax|
  \begin{myitemize}
    \bluedescription {{ degrees of freedom (variables) in the lattice and/or Wyckoff positions of the structure (DFT-relaxed geometry).}}
    \bluetype \verbWrapBlue|array of strings|
  \end{myitemize}
\item \verbWrapBlue|aflow_prototype_params_values_relax|
  \begin{myitemize}
    \bluedescription {{ values specifying the degrees of freedom of the structure (DFT-relaxed geometry).}} 
    \bluetype \verbWrapBlue|array of floats|
  \end{myitemize}
\end{myitemize}

%%%%%%%%%%%%%%%%%%%%%%%%%%%%%%%%%%%%%%%%%%%%%%%%%%%%%%%%%%%%%%%%%%%%% 
%% Data availability
%%%%%%%%%%%%%%%%%%%%%%%%%%%%%%%%%%%%%%%%%%%%%%%%%%%%%%%%%%%%%%%%%%%%%
\ \\

All crystallographic structure data is freely available and accessible online through \AFLOW.org or programmatically via 
the \REST- and \AFLUX\ Search-\APIS.
The \AFLOW\ prototype information is provided online at 
\href{http://aflow.org/prototype-encyclopedia}{http://aflow.org/prototype-encyclopedia}, 
and the corresponding structures can be generated with the \AFLOW\ source code.

%%%%%%%%%%%%%%%%%%%%%%%%%%%%%%%%%%%%%%%%%%%%%%%%%%%%%%%%%%%%%%%%%%%%% 
%% Code availability
%%%%%%%%%%%%%%%%%%%%%%%%%%%%%%%%%%%%%%%%%%%%%%%%%%%%%%%%%%%%%%%%%%%%% 
The \AFLOWXTALFINDERSHORT\ module is integrated into the \AFLOW\ software 
(version \AFLOWXTALFINDERAFLOWVERSION\ and later).
The source code for \AFLOW\ is available at
\href{http://aflow.org/install-aflow/}{http://aflow.org/install-aflow/}
and
\href{http://materials.duke.edu/AFLOW/}{http://materials.duke.edu/AFLOW/},
and it is compatible with most Linux, macOS, and Microsoft operating systems.
The multithreaded capabilities require GNU g++-4.4 or later.

%%%%%%%%%%%%%%%%%%%%%%%%%%%%%%%%%%%%%%%%%%%%%%%%%%%%%%%%%%%%%%%%%%%%% 
%% Acknowledgments
%%%%%%%%%%%%%%%%%%%%%%%%%%%%%%%%%%%%%%%%%%%%%%%%%%%%%%%%%%%%%%%%%%%%% 

%%%%%%%%%%%%%%%%%%%%%%%%%%%%%%%%%%%%%%%%%%%%%%%%%%%%%%%%%%%%%%%%%%%%% 
%% Author contributions
%%%%%%%%%%%%%%%%%%%%%%%%%%%%%%%%%%%%%%%%%%%%%%%%%%%%%%%%%%%%%%%%%%%%% 

%%%%%%%%%%%%%%%%%%%%%%%%%%%%%%%%%%%%%%%%%%%%%%%%%%%%%%%%%%%%%%%%%%%%% 
%% Additional Information 
%%%%%%%%%%%%%%%%%%%%%%%%%%%%%%%%%%%%%%%%%%%%%%%%%%%%%%%%%%%%%%%%%%%%% 
\noindent\textbf{Supplementary information.} The article is accompanied by supplementary information providing 
\textbf{i.} the computational details for the data in Figure~\ref{fig6_label} and 
\textbf{ii.} the full prototype list extracted from the \AFLOW-\ICSD\ catalog (a continuation of Table~\ref{table:common_prototypes}).

\noindent\textbf{\AFLOWXTALFINDER\ support.} Questions and bug reports should be emailed to \texttt{aflow@groups.io} 
with a subject line containing ``\AFLOWXTALFINDERSHORT''.

\newcommand{\Ozolins}{Ozoli{\c{n}}{\v{s}}}

\end{document}